\documentclass[journal,10pt,onecolumn,draftclsnofoot]{IEEEtran}
\usepackage{verbatim}
\usepackage{hyperref}
\usepackage[caption=false,font=footnotesize]{subfig}
\usepackage[dvips]{graphicx}
\usepackage{enumitem}
\usepackage{amsmath,amssymb}
\usepackage{amsthm}
\usepackage{cite}
\usepackage{stfloats}
\theoremstyle{plain}
\newtheorem{theorem}{Theorem}
\newtheorem{lemma}{Lemma}

\newtheorem{corollary}{Corollary}
\theoremstyle{definition}
\newtheorem{definition}{Definition}
\newtheorem{example}{Example}
\usepackage{enumitem}
\theoremstyle{remark}

\newtheorem{case}{Case}
\newcommand{\floor}[1]{\lfloor #1 \rfloor}
\newcommand{\ceil}[1]{\lceil #1 \rceil}
\usepackage{array}
\usepackage{tabularx}
\usepackage{ragged2e} 
\usepackage{url}
\allowdisplaybreaks
\title{Multi-Access Coded Caching Schemes from Maximal Cross Resolvable Designs}
\author{Niladri Das and
        B. Sundar Rajan, \textit{Fellow}, \textit{IEEE}% <-this % stops a space
\thanks{Manuscript received on date xx; revised on date xx; accepted on date xx. Date of publication xx,; date of current version xx. The editor coordinating the review of this paper and approving it for publication was Prof. xx.}
\thanks{Niladri Das and B. Sundar Rajan are with the Department
of Electrical Communication Engineering, Indian Institute of Science, Bangalore, 
India e-mail: niladridas@iisc.ac.in and bsrajan@iisc.ac.in.}% <-this % stops a space
\thanks{This work was supported partly by the Science and Engineering Research Board (SERB) of Department of Science and Technology (DST), Government of India, through J.C. Bose National Fellowship to B. Sundar Rajan, and through SERB-National Post Doctoral Fellowship (No. PDF/2019/000044) to Niladri Das.}}
\begin{document}
\maketitle
%A decoding strategy must exploit the u2c bipartite graph.
\begin{abstract}
We study the problem of multi-access coded caching (MACC): a central server has $N$ files, $K$ ($K \leq N$) caches each of which stores $M$ out of the $N$ files, $K$ users each of which demands one out of the $N$ files, and each user accesses $z$ caches. The objective is to jointly design the placement, delivery, and user-to-cache association, to optimize the achievable rate. This problem has been extensively studied in the literature under the assumption that a user accesses only one cache. However, when a user accesses more caches, this problem has been studied only under the assumption that a user accesses $z$ consecutive caches with a cyclic wrap-around over the boundaries. 
A natural question is how other user-to-cache associations fare against the cyclic wrap-around user-to-cache association. A bipartite graph can describe a general user-to-cache association. We identify a class of bipartite graphs that, when used as a user-to-cache association, achieves either a lesser rate or a lesser subpacketization than all other existing MACC schemes using a cyclic wrap-around user-to-cache association. The placement and delivery strategy of our MACC scheme is constructed using a combinatorial structure called maximal cross resolvable design.  
%
%which we show there exists a MACC scheme. And, 
%
% For a large class of user-to-cache association we provide a solution to the MACC problem. We show that for each 
%
%In this paper, if the bipartite graph depicting the user-to-cache association of a MACC problem satisfies three graph-theoretic conditions, we provide a solution to the MACC problem.

%The placement and delivery strategy is guided by a combinatorial structure called maximal cross resolvable design. 
% For MCRD based placements, we introduce a concept of called demand graph. The delivery strategy is based on certain matchings of the demand graph. We also identify a link between matchings of demand graphs and cliques of side information graphs.

%newly removed
%The problem is to reduce the number of files the server broadcasts to meet the users' demands. 

%We end with a discussion on cliques enforced by matchings of demand graphs and its applicability.
%We end with a discussion on the similarities between matchings of demand graphs and cliques of side information graphs.
%We conclude the paper with a discussion on a possible connection between number of edges in a matching and coding gain.
%with an example where the exists
%
%with a discussion on possible ways to further extend our work and incorporate more general user-to-cache associations.
%
%,   Our scheme accommodates all  whose corresponding . 
\end{abstract}

%all new lines are prefixed by "new line here"
\section{Introduction}
Consider the following scenario. A cellular cell is deployed with a base station (BS), and several access points (APs) are distributed within the cell. APs typically have a small coverage area than BS, cater to fewer users than BS, and may have a wire-line connection to the internet backbone. Each AP is served with memory and acts as a cache. This paper uses the word AP and cache alternatively, indicating the same device functionality.

A large number of files are stored with the BS. There are users in the cell connected (wireless) to the BS. A user may request to download files from the BS. Each user is also connected to one or more APs. The APs fetch data (before knowing the users' demands) from the BSs when internet traffic is low -- this is called the cache placement phase. This cache filling strategy is centralized, meaning that the BS decides which cache stores which parts of the files. Furthermore, the cache filling strategy is uncoded, meaning that the data placed at the APs are copies of segments of the files present at the BS.
%, or it could be decentralized -- where the APs arbitrarily select the bits of files to cache

User-to-cache association is a list that specifies which users access which caches. The central server may decide the user-to-cache association. The users inform of their demands (a demand is a request for a single file) to the BS. Based on a predetermined delivery strategy, the BS broadcasts a set of files to all users to fulfill their demands of the users. The term delivery load quantifies the number of files broadcast by the BS. The goal is to design the placement and the delivery policy such that the least pressure is put on the internet traffic during peak hours, which amounts to minimizing the number of files broadcast during peak hours. We assume that very high-speed links connect a user to the APs it accesses. Hence, the burden on the internet traffic is due to the files broadcast by the BS. Furthermore, we assume that the number of files is much more than the number of users. Hence, it is beneficial to minimize the delivery load assuming all users demand distinct files.%, which is the worst-case scenario..%, which is the worst-case scenario.

%new line here yctc
From the works conducted in the literature towards solving this problem, we see a trade-off between the delivery load and the size of the files. As in, a delivery load may be achievable only if the size of each file is greater than a certain value. Shanmugam \textit{et al.} in \cite{shanmugam} established a trade-off between the size of the files and the delivery load. Yan \textit{et al.} in \cite{yctc} also showed a trade-off between file size and delivery load, albeit for a particular placement. The size of the file is quantified by a term called subpacketization level. If a solution to the problem has a subpacketization level $F$, the solution requires each file to be split into $F$ subfiles. Furthermore, the higher the subpacketization level, the more computations are necessary to decode at the users. So it is also beneficial to minimize the subpacketization level. 

% must have at least $F$ bits. Furthermore, 

Traditionally, the content has been cached based on the file's popularity. Maddah-Ali and Niesen, in the seminal paper \cite{ali}, showed that by jointly designing the cache placement phase and the file delivery phase, an additional gain (coding gain) can be achieved. The work of \cite{ali} significantly improved the previous caching schemes (which may be referred to as caching schemes with uncoded delivery). Maddah-Ali \textit{et al.} also showed a trade-off between the memory size of the caches and the achievable rate. Furthermore, they showed that the rate achieved by their scheme is within $12$ times the information-theoretic optimum value. The same authors provided a decentralized solution to the coded caching problem in \cite{ali2}. 

%MAN scheme splits each file into a number of subfiles. This number is known as the subpacketization level.

Maddah-Ali and Niesen's caching model inherently assumes that each user accesses only one cache. We call such schemes as dedicated cache coded coding schemes. %; in literature they have been generally refered as dedicated cache coding schemes. 
If a user can connect to several APs available in its vicinity, then restricting the user to connect to only one AP may not be the best choice. Since granting access to multiple APs does not require extra hardware, such a scheme may provide better performance at the cost of more signaling. 
% solution while only making protocol-level changes.

Hachem \textit{et al.} in \cite{hachem} improvised the model (considered in \cite{ali}) to include scenarios where a user accesses more than one cache. They called this problem as the multi-access coded caching (MACC) problem. In their model, there are $K$ caches and $K$ users, each user accesses $z$ consecutive caches with a cyclic wrap-around over the boundaries, and each cache is accessed by $z$ caches. We call such a user-to-cache association a cyclic wrap-around user-to-cache association. Hachem \textit{et al.} showed a MACC scheme for any cache memory size $M$ files and access degree $z$.

%We split the works published on MACC after the work of Hachem \textit{et al.} into two categories: the first category is where it is

Subsequently, all works on MACC (which considers the number of users to be the same as the number of caches) assume a cyclic wrap-around user-to-cache association. However, reference \cite{hachem} quotes the following: ``we assume also for simplicity that the caches are arranged linearly and that users connect to $d_i$ consecutive caches, with a cyclic wrap-around for symmetry.'' Although we acknowledge that studying a simpler version of a problem can provide significant insights, until now, to the best of our knowledge, no work has justified the assumption of cyclic wrap-around user-to-cache association with any reason other than simplicity.

A bipartite graph can represent a general user-to-cache association of the MACC problem. It is an open problem to find out which bipartite graph provides the best performance when used as a user-to-cache association. The criteria for best performance may vary with applications--most applications would likely desire to achieve the least delivery load possible for a given upper limit on the subpacketization level.

 %Since there is a trade-off between delivery load and subpacketization level (the trade-off not completely established), 

%It is a natural question that whether there exist other user-to-cache associations which can provide better results than the cyclic wrap-around user-to-cache association results in the best performance. 

\subsection{State of the art}
Hachem \textit{et al.} were the first to study the MACC problem \cite{hachem}. They studied the problem assuming a cyclic wrap-around user-to-cache association. Subsequent to the work of Hachem \textit{et al.}, the MACC problem under the assumption of a cyclic wrap-around user-to-cache association has been studied by the following authors in the respective references: Serbetci \textit{et al.} in \cite{serbetci},   Reddy \textit{et al.} in \cite{reddy} and \cite{reddy2}, Cheng \textit{et al.} in \cite{cheng}, Sasi \textit{et al.} in \cite{shanuja} and \cite{shanuja2}, Mahesh \textit{et al.} in \cite{anjana}.

There have been some works in the literature that studied a version of the MACC problem where the number of users is more than the number of caches. %The rest of the introduction provides a lists of such works. multi-accesss coded caching
Katyal \textit{et al.} in \cite{digvijay} studied a version of a MACC problem where the user-to-cache association is decided by combinatorial structure called cross resolvable design (CRD). Using the CRD, they constructed a MACC scheme for this problem. The authors identified two infinite classes of CRDs from the existing literature in combinatorial designs and showed the performance of the MACC schemes constructed using these CRDs. Subsequent to \cite{digvijay}, Muralidhar \textit{et al.} in \cite{gen_digvijay} showed a generalization of the scheme in \cite{digvijay}. In reference \cite{gen_crd}, Muralidhar \textit{et al.} showed a new class of CRDs and the corresponding MACC schemes constructed using these CRDs.

%the MACC problem under a set up different to \cite{hachem}. They study a MACC problem where the number of users may be more than the number of caches. They construct a MACC scheme by using a combinatorial structure called cross resolvable design (CRD). CRDs are studied in a branch of mathematics called combinatorial design theory.  Given a CRD, the user-to-cache association, placement, and delivery of the MACC problem studied in \cite{digvijay} is decided by the CRD. T

% Katyal \textit{et al.} also introduced the notion of rate per user, which is equal to rate divided by the number of users. 

Muralidhar \textit{et al.} in \cite{pooja} studied a MACC problem where the user-to-cache association mimics the combination network, which is well-studied in the network coding literature. They showed that a generalization of the MAN scheme provides a solution to this problem. Recently Brunero \textit{et al.} in \cite{brunero} studied a generalization of the user-to-cache association considered by \cite{pooja}. They showed a solution and proved that it achieves the optimal delivery load under uncoded placement. The authors of \cite{brunero} also address a scenario where the server is unaware of the user-to-cache association during cache placement.

It is to be noted that the works of \cite{digvijay,gen_digvijay,gen_crd,pooja,brunero} do not intersect with our work except for the trivial cases, which is when either each user accesses only one cache or each user accesses all caches. For the rest of this document, we limit ourselves to MACC problems where the number of users is the same as caches. % as it was defined originally in \cite{hachem}.

%the words ``the MACC problem'' refer to the  problem where the number of users are the same as the number of caches. 

%  extend the MAN scheme to obtain a MACC scheme. 

% discuss our core observation that guided us to construct our scheme. In Section~\ref{conclusion} we conclude the paper.

%In Section~\ref{discuss} we discuss our core observation that guided us to construct our scheme. 

\subsection{Contributions}\label{contributions}
%The salient contributions of this paper are as follows. 
\begin{enumerate}
\item We identify a class of bipartite graphs. If the server chooses the user-to-cache association (which can be represented by a bipartite graph) of the MACC problem from this class of bipartite graphs, we show a MACC scheme that provides a solution to the MACC problem. This contribution is shown in Theorem~\ref{theorem1} of Section~\ref{main}. 

\item We show that our MACC scheme achieves a better trade-off between the achievable rate and subpacketization level compared to all other existing MACC schemes, at least for some memory sizes. The comparison is shown in Section~\ref{compare}. 

\item The placement and delivery strategy of our MACC scheme is based on a combinatorial design called maximal cross resolvable design (MCRD). In Theorem~\ref{theorem-1} of Section~\ref{sec2} we show a construction of MCRDs with general parameters.

\end{enumerate}

\subsection{Organization of the paper}
%Write afresh
%In Section~\ref{contributions} we  highlight the contributions of the paper. 
In Section~\ref{problem} we state the system model. In Section~\ref{preli:g} we reproduce some definitions and lemmas from the literature on graph theory. In Section~\ref{sec2} we first reproduce the definition of MCRD from the literature on combinatorial design theory, and then show a construction of MCRDs with general parameters. In Section~\ref{sec3a} we isolate a class of user-to-cache association bipartite graphs. In Section~\ref{main} we show our MACC scheme. In Section~\ref{compare} we compare our scheme with other relevant MACC schemes. In Section~\ref{example} we provide two examples of the placement and delivery strategy of our MACC scheme. The paper is concluded in Section~\ref{conclusion}.

%In Section~\ref{conclusion} we conclude the paper.
% that produces a solution of the MACC problem whose user-to-cache association belongs to this class of graphs
% In Section~\ref{proof-here} we prove a Theorem we stated in Section~\ref{sec3a}.
\section{System Model}\label{problem}
%In this section, we give a formal description of the system model. 
There is a central server, $K$ caches, and $K$ users. Let $U = \{k_1,k_2,\ldots,k_K\}$ be the set of the users, and $C = \{c_1,c_2,\ldots,c_K\}$ be the set of caches. So we have $|U| = |C| = K$. 
\begin{definition}
A user-to-cache association bipartite graph is a bipartite graph $G = (U,C,E)$ where $U \cup C$ is the vertex set of $G$, sets $U$ and $C$ are disjoint, every edge in $E$ connects one vertex in $U$ to another vertex in $C$, an edge connects $u \in U$ to $v \in C$ if and only if user $u$ accesses cache $c$.
\end{definition}
The central server decides a user-to-cache association bipartite graph such that each user accesses $z \geq 1$ caches. Here $z$ is called the access degree.

The central server contains $N$ files $W^1,W^2,\ldots,W^N$. Each file $W^i$ is split into $F$ subfiles $W^i(1), W^i(2), \ldots, W^i(F)$. Here $F$ is called the subpacketization level. The number $j$ in $W^i(j)$ is called as the index of the subfile $W^i(j)$. Each subfile contains $s$ bits. Each of the $K$ caches can store $MF$ subfiles in its memory. The ratio $M/N$ is called normalized memory size. It is assumed that each user can instantaneously download the contents of the caches it accesses.

We describe a $(K,z,M,R)$ multi-access coded caching scheme with uncoded placement. There are two phases: the placement phase and the delivery phase. The placement phase comes first. During the placement phase, the cache memory is filled up without knowing the users' future demands. Each cache $c_i$ for $1\leq i\leq K$ stores $MF$ subfiles out of the $NF$ subfiles available with the server. That is, if $Z_i$ denote the content of the cache $c_i$, then $Z_i \subseteq \{W^1(1), W^1(2), \ldots, W^N(F)\}$ and $|Z_i| = MF$. 

% uses a function (called caching function) 
%\begin{equation}
%f_{c_i}: 2^{sNF} \to 2^{sMF}
%\end{equation}
%such that $f_{c_i}()$
%
%to store $MF$ subfiles its cache. 

During the delivery phase, each user makes a demand of a single file. Say the user $k_i$ for $1\leq i\leq K$ demands the file $W^{d_{k_i}}$ where $1\leq d_{k_i}\leq N$. On knowing the demands of all users, the server uses a function (called encoding function) 
\begin{equation}
f_{(d_{k_1},d_{k_2},\ldots,d_{k_K})}: 2^{sNF}   \to 2^{sRF}
\end{equation}
to broadcast $R$ files. Here $R$ is called the (achievable) rate or the delivery load. % Say for $1\leq i\leq K$ user $k_i$ accesses the caches in the set $C_{k_i}$ where $C_{k_i} \subseteq \{1,2,\ldots,K\}$ and $|C_{k_i}| = z$.  
Each user $k_i$ for $1\leq i\leq K$ uses a function (called decoding function) 
\begin{equation}
f_{(k_i, d_{k_1},d_{k_2},\ldots,d_{k_K})}: 2^{sRF} \times 2^{zsMF} \to 2^{sF}
\end{equation}
in an attempt to retrieve its demanded file of $F$ subfiles from the $RF$ subfiles broadcast by the server and the memory contents of the $z$ caches that user $k_i$ accesses. %This completes the description of a $(K,z,M,R)$ MACC scheme with uncoded placement.

Suppose each user can successfully retrieve its demanded file using the $R$ files broadcast by the server and the memory contents of the $z$ caches it accesses. In that case, the placement and delivery policy and the user-to-cache association together is called a $(K,z,M,R)$ MACC scheme. The objective is to jointly optimize the achievable rate ($R$ files) and the subpacketization level by designing the placement and delivery policy and the user-to-cache association. The coding gain of a $(K,z,M,R)$ MACC scheme is the number of users who benefit from each transmission made by the server during the delivery phase.

%If each user can successfully retrieve its demanded file using a $(K,z,M,R)$ MACC scheme, the scheme is said to be a solution to the MACC problem. The objective is to find a solution which jointly optimizes the achievable rate ($R$) and the subpacketization level. The coding gain of a $(K,z,M,R)$ MACC scheme is the number of users who benefit from each transmission made by the server during the delivery phase.

% $(K,z,M,R)$ MACC scheme with the least achievable rate ($R$), keeping the subpacketization level within a permissible upper limit. 

% The objective is to find $(K,z,M,R)$ MACC schemes that acts as a solution for a given user-to-cache association bipartite graph. If there are many solutions we want the solution with least achievable rate ($R$) and least subpacketization level. 

\section{Preliminaries on graph theory}\label{preli:g}
%The preliminaries consist of some definitions from graph theory and some definitions from combinatorial design theory. 
An edge $e$ of a graph is said to connect vertex $u$ to vertex $v$ if and only if one end point of the edge is $u$ and the other edge point of the edge is $v$. Equivalently, one can say that the edge $e$ is adjacent to vertices $u$ and $v$. %We defer the latter definitions to Section~\ref{sec2}.
%\subsection{Preliminaries on graph theory}
\begin{definition}
Bipartite graph: A bipartite graph $G = (V_1,V_2,E)$ is such that $V_1 \cap V_2 = \emptyset$, and there does not exist any edge that is adjacent to two vertices in $V_i$ for $i = 1,2$.
%no edge is adjacent to two vertices from $V_1$ or two vertices from $V_2$. 
\end{definition}

\begin{definition}
Matching of a bipartite graph $G = (V_1,V_2,E)$: a matching is a subset $E^\prime \subseteq E$ such that if $e_1, e_2 \in E$ then there exists no vertex $v \in V_1 \cup V_2$ such that both $e_1$ and $e_2$ are adjacent to $v$. The size of a matching is the number of edges in the matching.
\end{definition}

\begin{definition}
Vertex cover of a matching of a bipartite graph $G = (V_1,V_2,E)$: let $E^\prime \subseteq E$ be a matching of the bipartite graph $G = (V_1,V_2,E)$. Vertex cover of $E^\prime$ is the set of all vertices that are adjacent to some edge in $E^\prime$. Vertex cover of $E^\prime$ is denoted by $\nu(E^\prime)$. 
\end{definition}

\begin{definition}
Partial matching of a bipartite graph $G = (V_1,V_2,E)$: let $E^\prime \subseteq E$ be a matching of the bipartite graph $G = (V_1,V_2,E)$. If $\nu(E^\prime) \subseteq V_1 \cup V_2$ then $E^\prime \subseteq E$ is a partial matching of $G = (V_1,V_2,E)$.
\end{definition}

\begin{definition}
Perfect matching of a bipartite graph $G = (V_1,V_2,E)$: let $E^\prime \subseteq E$ be a matching of the bipartite graph $G = (V_1,V_2,E)$. If $\nu(E^\prime) = V_1 \cup V_2$ then $E^\prime \subseteq E$ is a perfect matching of $G = (V_1,V_2,E)$.
\end{definition}

%\begin{definition}
%\textbf{Maximal matching of a bipartite graph $G = (V_1,V_2,E)$.} Say $E^\prime \subseteq E$ is a matching of the bipartite graph $G = (V_1,V_2,E)$. If there exists no $E^{\prime\prime}\subseteq E$ such that $E^\prime \subset E^{\prime\prime}$ and $E^{\prime\prime}$ is a matching of $G = (V_1,V_2,E)$, then $E^\prime$ is a maximal matching of $G = (V_1,V_2,E)$.
%\end{definition}

%\subsection{Preliminaries on Maximal Cross Resolvable Designs}\label{sec2}
\section{Maximal Cross Resolvable Designs}\label{sec2}
\begin{definition}
A design $(X,\mathcal{A})$ is an ordered tuple of two finite sets $X$ and $\mathcal{A} \in 2^X$. The elements of the set $X$ are called points. The elements of  the set $\mathcal{A}$ are called blocks. Each block in $\mathcal{A}$ is a subset of $X$. Any two blocks in $\mathcal{A}$ contain the same number of points. 
\end{definition}
\begin{example}\label{ex-a1}
Let $X = \{1,2,3,4,5,6,7,8,9,10,11,12,13,14,15,16\}$, and $\mathcal{A} = \{ \{1,2,3,4\},\allowbreak   \{1,5,6,7\},\allowbreak  \{1,8,9,10\},\allowbreak \{2,11,12,13\},\allowbreak \{3,14,15,16\},\allowbreak  \{4,5,6,7\},\allowbreak \{8,9,10,11\}, \allowbreak \{12,13,14,15\} \}$. Then $(X,\mathcal{A})$ is a design.
\end{example}
\begin{definition}
A parallel class $\mathcal{P}$ of a design $(X,\mathcal{A})$ is a subset of $\mathcal{A}$ such that $\mathcal{P}$ partitions $X$.
\end{definition}
\begin{example}\label{ex0}
In Example~\ref{ex-a1}, $\mathcal{P} =\{ \{1,8,9,10\}, \{2,11,12,13\}, \{3,14,15,16\},  \{4,5,6,7\} \}$ is a parallel class.
\end{example}
%
%Since any two blocks contains the same number of elements, any two parallel classes must contain the same number of blocks. 
Since any two blocks has the same number of elements, and each parallel class partitions $X$, any two parallel class must have the same number of blocks. We denote the number of blocks in a parallel class by $b$. In Example~\ref{ex0}, $b = 4$.
\begin{definition}
A design $(X,\mathcal{A})$ is a resolvable design if for some positive integer $m$, the design has $m$ parallel classes $\mathcal{P}_1, \mathcal{P}_2,\ldots, \mathcal{P}_m$ such that  $\mathcal{P}_1, \mathcal{P}_2,\ldots, \mathcal{P}_m$ partitions $\mathcal{A}$.
\end{definition}
\begin{example}
It can be seen that the design $(X,\mathcal{A})$ considered in Example~\ref{ex-a1} is not resolvable. 
\end{example}
\begin{example}\label{ex-a2}
Let $X = \{1,2,3,4,5,6,7,8,9,10,11,12,13,14,15,16\}$. And $\mathcal{A} = \{ \{1,2,3,4\}, \allowbreak  \{5,6,7,8\},\allowbreak  \{9,10,11,12\},\allowbreak \{13,14,15,16\},\allowbreak \{1,2,9,13\},\allowbreak  \{5,6,10,14\}, \{3,7,11,15\},\allowbreak   \{4,8,12,16\} \}$. Then $(X,\mathcal{A})$ is a design. Consider the two parallel classes of $(X,\mathcal{A})$.
\begin{IEEEeqnarray*}{l}
\mathcal{P}_1 = \{ \{1,2,3,4\},  \{5,6,7,8\},  \{9,10,11,12\}, \{13,14,15,16\} \}\\
\mathcal{P}_2 = \{ \{1,2,9,13\},  \{5,6,10,14\}, \{3,7,11,15\},  \{4,8,12,16\} \}.
\end{IEEEeqnarray*}
It can be seen that each sets $\mathcal{P}_1$ and $\mathcal{P}_2$ partition $X$, and they together partition $\mathcal{A}$. So $(X,\mathcal{A})$ is a resolvable design. For this design, $b = 4$.
\end{example}
\begin{definition}
A resolvable design $(X, \mathcal{A})$ with $m$ parallel classes is called a maximal cross resolvable design (MCRD) if for some positive integer $\mu_m$, any $m$ blocks chosen from $m$ distinct parallel classes have $\mu_m$ elements in common. That is, if for some ordering of the blocks in parallel class $\mathcal{P}_i$ the $j^{\text{th}}$ block is denoted by $B(i,j_i)$ for $1\leq i\leq m$, $1\leq j_i\leq b$,  then for any $1\leq j_i \leq b$, $|\cap_{i=1}^m B(i,j_i)| = \mu_m$. The integer $\mu_m$ is called $m^{\textit{th}}$ cross intersection number.
\end{definition}
\begin{example}
The resolvable design shown in Example~\ref{ex-a2} is not a maximal cross resolvable design as: between $B(1,1) = \{1,2,3,4\}$, $B(2,1) = \{1,2,9,13\}$ we have $|B(1,1) \cap B(2,1)| = 2$, but between $B(1,2) = \{5,6,7,8\}$ and $B(2,1) = \{1,2,9,13\}$ we have $B(1,2) \cap B(2,1) = 0$. 
\end{example}
\begin{example}\label{ex-a3}
Let $X = \{1,2,3,4,5,6,7,8,9,10,11,12,13,14,15,16\}$. And $\mathcal{A} = \{ \{1,2,3,4\},\allowbreak   \{5,6,7,8\},\allowbreak  \{9,10,11,12\},\allowbreak \{13,14,15,16\},\allowbreak \{1,5,9,13\},\allowbreak  \{2,6,10,14\}, \{3,7,11,15\},\allowbreak   \{4,8,12,6\} \}$. Then $(X,\mathcal{A})$ is a design. Consider the two parallel classes of $(X,\mathcal{A})$.
\begin{IEEEeqnarray*}{l}
\mathcal{P}_1 = \{ \{1,2,3,4\},  \{5,6,7,8\},  \{9,10,11,12\}, \{13,14,15,16\} \}\\
\mathcal{P}_2 = \{ \{1,5,9,13\},  \{2,6,10,14\}, \{3,7,11,15\},  \{4,8,12,16\} \}.
\end{IEEEeqnarray*}
It can be seen that each sets $\mathcal{P}_1$ and $\mathcal{P}_2$ partition $X$, and they together partition $\mathcal{A}$. Furthermore, the cardinality of the intersection between any two blocks selected from the two distinct parallel classes is always $1$. So the design is a maximal cross resolvable design (MCRD) with $\mu_2 = 1$.
\end{example}
%
%\begin{definition}
%A resolvable design $(X, \mathcal{A})$ with $m$ parallel classes is called a cross resolvable design (CRD) if for some positive integer $\mu_i$, any $i$ blocks chosen from any $i$ distinct parallel classes have $\mu_i$ elements in common. The integer $\mu_i$ is called $i^{\textit{th}}$ cross intersection number.
%\end{definition}
%It can be seen that every CRD is an MCRD if $\mu_m$ exists. 
%That is, if for some ordering of the blocks in parallel class $\mathcal{P_i}$ the $j^{\text{th}}$ block is denoted by $B(i,j) \in \mathcal{P}_i$ for $1\leq i\leq m$, $1\leq j_i\leq b$,  then for any $1\leq j_i \leq b$, $|\cap_{i=1}^m B(i,j_i)| = \mu_m$.
%
%Since any two blocks has the same number of elements, and each parallel class partitions $X$, any two parallel class must have the same number of blocks.
\begin{lemma}\label{lemmax1}
Let $(X, \mathcal{A})$ be an MCRD with $m$ parallel classes and $\mu_m = 1$. Furthermore, say each parallel class contains $b$ blocks. Then, each block has exactly $b^{m-1}$ elements. Moreover, $|X| = b^m$.
\end{lemma}
The proof of Lemma~\ref{lemmax1} is deferred till Appendix~\ref{proof:lemmax1}.

%
%The following Corollary~\ref{coro1} is evident from the proof of the above lemma. The delivery scheme of our MAAC scheme will be based on this corollary. 

%For $1\leq n\leq m$, $n\neq i$, $1\leq l_n \leq b$, 
We consider a function $I: \{1,2,\ldots,b\}^m \to X$ defined as $I(l_1,l_2,\ldots,l_m) = \cap_{n=1}^{m} B(n,l_n)$. The set \newline $\{l_1,l_2,\cdots,l_{i-1},l_{i+1},\cdots,l_m\}$ can be chosen in $b^{m-1}$ ways, and for each such choice there is one instance of the set $I(l_1,l_2,\ldots,l_{i-1},j,l_{i+1},\ldots,l_m) = \cap_{n=1,n\neq i}^{m} B(n,l_n) \cap B(i,j)$. 
\begin{corollary}\label{coro1}
Let $(X, \mathcal{A})$ be an MCRD with $m$ parallel classes. For $1\leq i\leq m$, $1\leq j\leq b$, say $B(i,j)$ is the $j^{\text{th}}$ block in the $i^{\text{th}}$ parallel class for some ordering of the parallel classes and the blocks in each parallel class. Then 
\begin{IEEEeqnarray}{l}
B(i,j) = \cup_{l_{1} = 1}^{b}  \cup_{l_{2} = 1}^{b} \cdots \cup_{l_{i-1} = 1}^{b} \cup_{l_{i+1} = 1}^{b}  \cdots \cup_{l_m = 1}^{b} I(l_1,l_2,\ldots,l_{i-1},j,l_{i+1},\ldots,l_m).
\end{IEEEeqnarray}
\end{corollary}
The proof of Corollary~\ref{coro1} is deferred to Appendix~\ref{proof:coro1}.

\begin{example}
In Example~\ref{ex-a3}, as per Lemma~\ref{lemmax1}, $m = 2$, $b = 4$, $\mu_m = 1$, and each block has $\mu_mb^{m-1} = 4$ elements. In congruence to Corollary~\ref{coro1}, we have:
\begin{IEEEeqnarray*}{l}
\cup_{l=1}^4 I(1,l) = \cup_{l=1}^4 B(1,1) \cap B(2,l) \\= \{ B(1,1) \cap B(2,1)\} \cup \{ B(1,1) \cap B(2,2)\} \cup \{ B(1,1) \cap B(2,3)\} \cup \{ B(1,1) \cap B(2,4)\}\\
= B(1,1) \cap \{B(2,1)\cup B(2,2) \cup B(2,3) \cup B(2,4) \} \\
= B(1,1) \cap X = B(1,1).
\end{IEEEeqnarray*}
\end{example}
\begin{example}\label{ex-a4}
Let $X = \{1,2,\ldots,27\}$. And $\mathcal{A}$ be the sets contained in the following parallel classes $\mathcal{P}_1$, $\mathcal{P}_2$, $\mathcal{P}_3$. 
\begin{IEEEeqnarray*}{l}
\mathcal{P}_1 = \{ \{1,2,3,4,5,6,7,8,9\},  \{10,11,12,13,14,15,16,17,18\},  \{19,20,21,22,23,24,25,26,27\}\}\\
\mathcal{P}_2 = \{ \{1,2,3,10,11,12,19,20,21\},  \{4,5,6,13,14,15,22,23,24\}, \{7,8,9,16,17,18,25,26,27\}\\
\mathcal{P}_3 = \{ \{1,4,7,10,13,16,19,22,25\},  \{2,5,8,11,14,17,20,23,26\}, \{3,6,9,12,15,18,21,24,27\}.
\end{IEEEeqnarray*}
It can be seen $(X,\mathcal{A})$ is a design with $b = 3$, $\mu_3 = 1$. Consistent with Lemma~\ref{lemmax1}, each block has $\mu_3 b^{2}$ elements. Furthermore, in congruence to Corollary~\ref{coro1} we have:
\begin{IEEEeqnarray*}{l}
\cup_{l_2 = 1}^3 \cup_{l_3 = 1}^3 I(1,l_2,l_3)\\ 
= \cup_{l_2 = 1}^3 \{I(1,l_2) \cap B(3,1)\} \cup  \{I(1,l_2) \cap B(3,1)\} \cup \{I(1,l_2) \cap B(3,1)\}\\
= \cup_{l_2 = 1}^3 I(1,l_2) \cap \{B(3,1)\cup B(3,2) \cup B(3,3)\}\\
= \cup_{l_2 = 1}^3 I(1,l_2) = \cup_{l_2 = 1}^3 \{B(1,1) \cap B(2,l_2)\}\\
= B(1,1) \cap \{B(2,1) \cup B(2,2) \cup B(2,3) \} = B(1,1).
\end{IEEEeqnarray*}
\end{example}

\begin{theorem}\label{theorem-1}
For any positive integer $m$, $n$, and $b$, there exists an MCRD $(X,\mathcal{A})$ with $X = \{1,2,\ldots,nb^m\}$, $|\mathcal{A}| = mb$, $m$ parallel classes, $b$ blocks in each parallel class, and $\mu_m = n$.
\end{theorem}
The proof of Theorem~\ref{theorem-1} is shown in Appendix~\ref{sysMCRD}.

%It can be seen that if there exists one solution, there would exist many other solutions. We want a solution with 

%A $(K,z,M,R)$ multi-access coded caching scheme is the ensemble of the $K$ caching functions, $N^K$ encoding functions, $KN^K$ decoding functions. 

%A $(K,z,M,R_u)$ MACC scheme is said to have uncoded delivery if the encoding function is such that the $R_uF$ subfiles broadcast by the server is a subset of the subfiles of $W^1,W^2,\ldots,W^N$. %This scheme is called optimum MACC scheme with uncoded delivery if there exists no $(K,z,M,R_u^*)$ MACC scheme such that $R_u^* < R_u$.

%If for a MACC problem there exists a $(K,z,M,R)$ scheme as well as a $(K,z,M,R_u)$ MACC scheme with uncoded delivery, then the ratio $R_u/R$ is called as coding gain of the $(K,z,M,R)$ MACC scheme over $(K,z,M,R_u)$ MACC scheme. The \textit{supremum} of $\{R_u/R \, | \, \text{ s. t. there exists a }  (K,z,M,R_u) \text{ MACC scheme with uncoded delivery} \}$ is called as the coding gain.
%

\section{A class of user-to-cache association bipartite graphs}\label{sec3a}
%
%We first describe the user-to-cache association for which our scheme is applicable.
%\subsection{User-to-cache association}\label{bipartite}
%%$•$
%
%Our scheme provides a solution to the MACC problem described in Section~\ref{problem} if for some integers $m$ and $b$ where $K = mb$ the user-to-cache association bipartite graph $G = (U,C,E)$ follows the following three conditions \textbf{C1, C2, C3}. 
Let $m$ and $b$ be integers where $K = mb$. Consider the class of all user-to-cache association bipartite graphs $G = (U,C,E)$ that follows the following three conditions \textbf{C1, C2, C3}. 
\begin{itemize}
\item Condition\textbf{ C1:} $G = (U,C,E)$ is a disjoint union of $m$ user-to-cache association bipartite graphs $G_i = (K_i,C_i,E_i)$ for $1\leq i\leq m$ such that $|K_i| = |C_i| = b$, if $j \neq i$, $K_i \cap K_j = \emptyset$, $C_i \cap C_j = \emptyset$.
\item Condition\textbf{ C2:} For $1\leq i\leq m$ the vertices $C_i$ can be partitioned into $z$ disjoint subsets $C_{(i,1)},C_{(i,2)},\ldots,C_{(i,z)}$ such that (i) for $1\leq l\leq z-1$ the subset $C_{(i,l)}$ contains $\floor{\frac{b}{z}}$ vertices, the subset $C_{(i,z)}$ contains $b - (z-1)\floor{\frac{b}{z}}$ vertices; (ii) any vertex $u \in K_i$ connects to at most one vertex in $C_{(i,l)}$ for $1\leq l\leq z$.
\item Condition\textbf{ C3:} There is a perfect matching in $G_i$ for $1\leq i\leq m$.
\end{itemize}

\begin{figure}[h]
\centering
\includegraphics[width=0.9\textwidth]{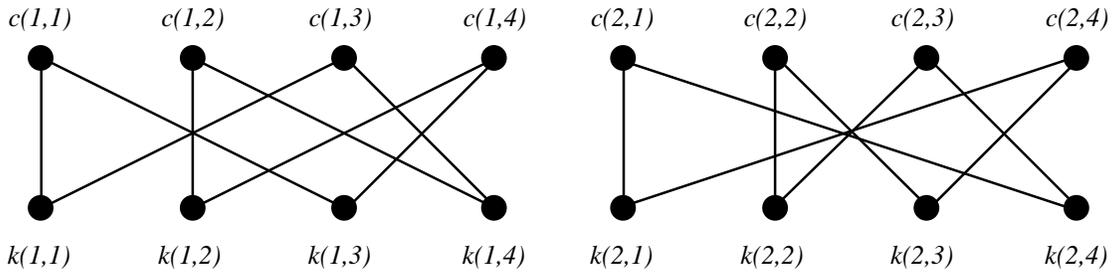}
\caption{An example user-to-cache association bipartite graph $G = (U,C,E)$ that satisfies conditions \textbf{C1, C2, C3}. There are $8$ users $U = \{k(i,j) \,| \, 1\leq i\leq 2, 1\leq j\leq 4 \}$, $8$ caches $C = \{c(i,j) \,| \, 1\leq i\leq 2, 1\leq j\leq 4 \}$, $b = 4$, $m=2$, and we have $z = 2$. A vertex labeled by $k(i,j)$ is connected to a vertex labeled by $c(i,j)$ if and only if user $k(i,j)$ accesses cache $c(i,j)$. For instance, user $k(1,1)$ accesses caches $c(1,1)$ and $c(1,3)$.}
\label{graph1}
\end{figure}
%\end{example}
%
Figure~\ref{graph1} shows an example user-to-cache association bipartite graph that satisfies conditions \textbf{C1, C2, C3}.

As a result of condition $\textbf{C1}$, a user $k\in K_i$ cannot access any caches from $C_j$ for $j\neq i$. As it will be shown in Section~\ref{delivery}, condition \textbf{C1} enforces a matching of size $m$, which in turn allows our scheme to achieve a coding gain of $m$: in our delivery algorithm, each transmission from the central server will be useful to one user in each $K_i$ for $1\leq i\leq m$. % we further elaborate this statement in Section~\ref{delivery}.
% after proving the correctness of our delivery scheme in Remark~\ref{intuition}.
% $m$ users with o

Let the $j^{\text{th}}$ user for $1\leq j\leq b$ in $K_i$ be denoted by $k(i,j)$ (for some ordering of the users in $K_i$). So  $K_i = \{k(i,1),k(i,2),\ldots,k(i,b)\}$. Similarly, let the $j^{\text{th}}$ cache for $1\leq j\leq b$ in $C_i$ be denoted by $c(i,j)$ (for some ordering of the caches in $C_i$). So $C_i = \{c(i,1),c(i,2),\ldots,c(i,b)\}$.

Conditions \textbf{C2, C3} apply on $G_i = (K_i,C_i,E_i)$ for $1\leq i\leq m$. Condition \textbf{C2} ensures that each user accesses at most $z$ caches. Furthermore, a user $k(i,j)$ can access at most one cache in $C_{(i,l)}$ for $1\leq l\leq z$. Specifically,  if $(k(i,j),c(i,j_1)), (k(i,j),c(i,j_2)) \in E_i$ and $c(i,j_1) \in C_{(i,n_1)}$, $c(i,j_2) \in C_{(i,n_2)}$, then $C_{(i,n_1)} \neq C_{(i,n_2)}$. Note, caches contained in $C_{(i,z)}$ are the caches that are in $C_i$ but not in the sets $C_{(i,l)}$ for $1\leq l\leq z-1$. The intuitive reasoning behind such partitioning is as follows. In the placement phase, we will place contents such a way that caches belonging to the separate partitions will never have any files in common; and caches belonging to the same partition may have files in common. Condition \textbf{C2} helps maximize the local coding gain.

Condition \textbf{C3} ensures that there is some set $M_i \subseteq E_i$ such that for each user $k(i,j)$ there is exactly one cache $c(i,j^\prime)$ such that $(k(i,j),c(i,j^\prime)) \in M_i$, and for each cache $c(i,l)$ there is only one user $k(i,l^\prime)$ such that $(k(i,l^\prime),c(i,l)) \in M_i$. So, condition \textbf{C3} ensures that there is a bijection $f_{M_i}$ from $K_i$ to $C_i$: if $(u,v) \in M_i$, $u$ maps to $v$.% For $1\leq i\leq m$, $1\leq j\leq b$, let $f_{M_i}^{-1}(c(i,j)) = k(i,j^*)$. We will use this function in our delivery algorithm.

%In Section~\ref{bipartite2} we further put forward our reasoning behind enforcing these conditions. %
%

% on the user-to-cache association bipartite graph.
%In this section we prove the following theorem.
%We prove the following theorem in this section.

%\subsection{MACC scheme}
\section{Main result and comparisons}\label{main}
\begin{theorem}\label{theorem1}
Consider a MACC problem with $K$ users, $K$ caches, access degree $z$, such that for some integers $m$ and $b$ where $K = mb$ the user-to-cache association bipartite graph satisfies the conditions \textbf{C1, C2, C3}. Let the central server have $N$ files where $N \geq K$, and the cache memory size be $M$ files. For $M = \frac{tN}{b}$ ($t \in \mathbb{N}$) and a subpacketization level $b^m$ a rate of $R$ files is achievable where
%There exists a solution to the MACC problem that uses a subpacketization level $b^m$ and for $M = \frac{tN}{b}$ achieves a rate of $R$ files where
\begin{IEEEeqnarray*}{l}
R = (b - tz) \;\text{ for }\; 1 \leq t \leq \floor{\frac{b}{z}}\\
R = (b - (z-1)\floor{\frac{b}{z}} - t) \; \text{ for }\; \floor{\frac{b}{z}} < t < b - (z-1)\floor{\frac{b}{z}}\\
R = 0 \;\text{ for }\; t \geq b - (z-1)\floor{\frac{b}{z}}.
\end{IEEEeqnarray*}
\end{theorem}
The proof of Theorem~\ref{theorem1} shown in Appendix~\ref{proof-here}.

\subsection{Rate-subpacketization trade-off}
It can be seen from Theorem~\ref{theorem1} that the achievable rate increases with $b$. Since $K = mb$, increasing $m$ decreases $b$, and thereby the achievable rate decreases with increasing $m$. On the other hand, the subpacketization level increases exponentially with $m$ and increases polynomially with $b$. So our scheme has a trade-off: reducing rate leads to a higher subpacketization, and reducing subpacketization leads to a higher rate. It is to be noted that our scheme requires $z \leq b$, and so decreasing $b$ limits the value of $z$.

\subsection{Comparison}\label{compare}
In this subsection, we compare the rate and subpacketization level of our scheme with the MACC schemes shown in references \cite{reddy2,reddy,anjana,shanuja,shanuja2,cheng,hachem,serbetci}, and show that at least for some memory sizes our scheme achieves either a lesser rate or a lesser subpacketization.

The scheme shown in \cite{serbetci} is denoted by SPE scheme, the scheme in \cite{reddy} is denoted by RK scheme, the scheme in \cite{cheng} is denoted by NT scheme, the scheme in \cite{reddy2} is denoted by SICPS scheme, the scheme in \cite{shanuja} is denoted by SR1 scheme, the scheme in \cite{shanuja2} is denoted by SR2 scheme, and the scheme in \cite{anjana} is denoted by MR scheme. 

\begin{figure}[!htpb]
\centering
%\ContinuedFloat 
\subfloat[]{\includegraphics[width=0.8\textwidth]{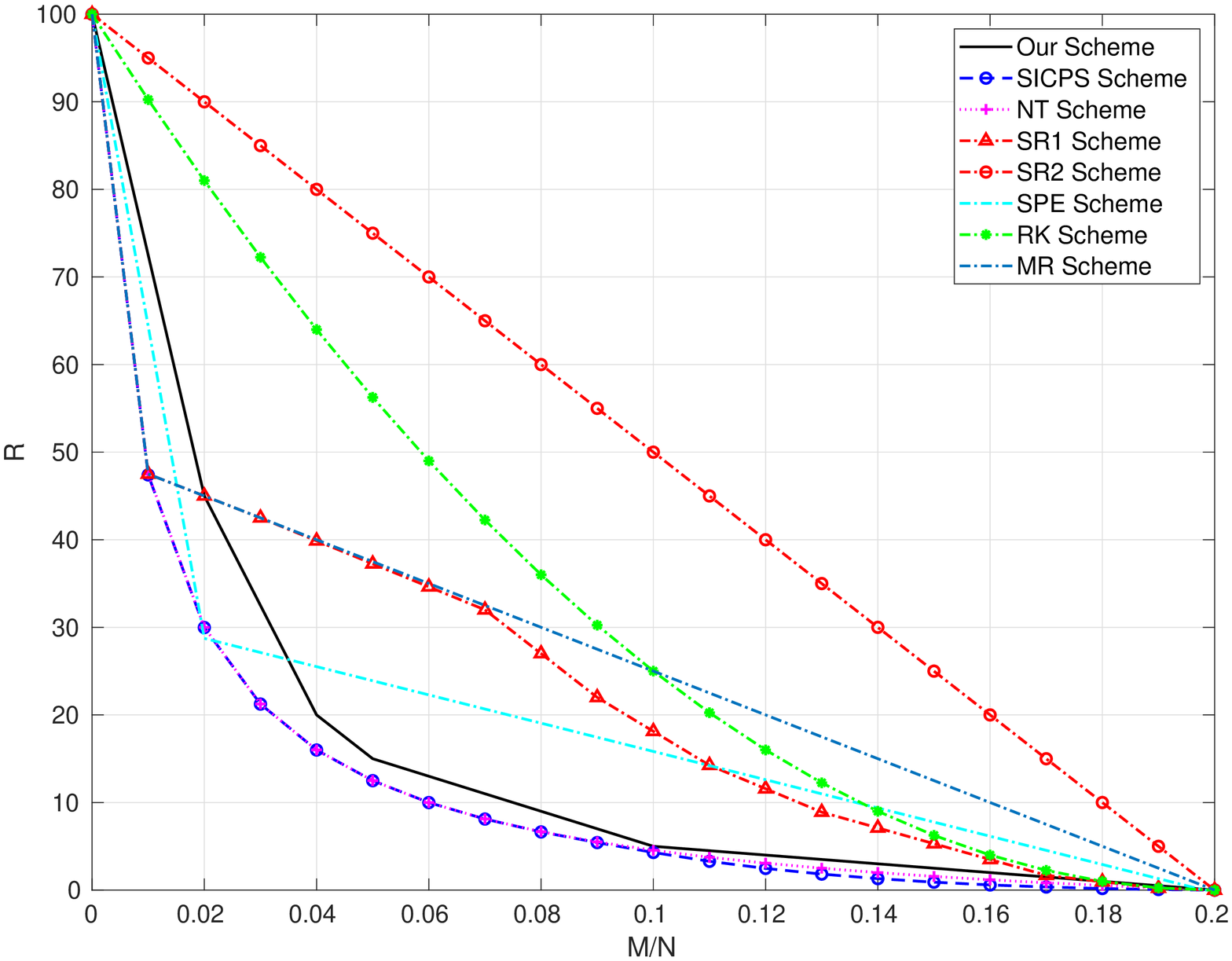}
\label{ratevsm}}
\hfil
%\vspace{-5pt}
\subfloat[]{\includegraphics[width=0.8\textwidth]{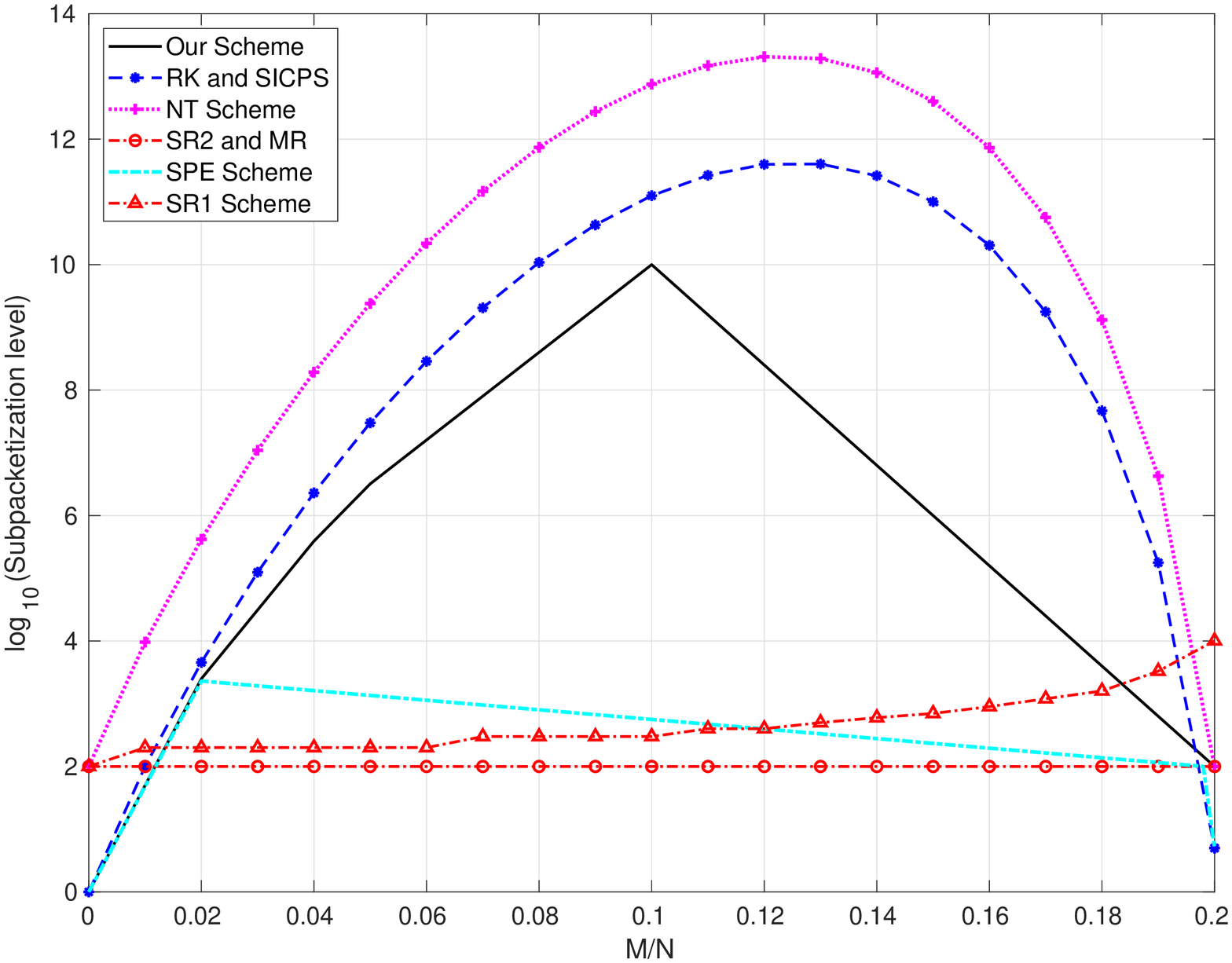}
\label{subvsm}}
\caption{Comparison of rate and subpacketization level of Our Scheme, the RK scheme, the SICPS scheme, the NT scheme, the SPE scheme, the SR1 scheme, the SR2 scheme, and the MR scheme for MACC problem with $K = 100$, $z = 5$. The subpacketization level in Fig.~\ref{subvsm} is shown in $log_{10}$ scale. It is to be noted that the achievable points in Fig.~\ref{subvsm} are connected by a line for aesthetics.} %Fig.~\ref{fig01} shows the curves for each of the individual schemes. Fig.~\ref{fig02} shows the same set of curves along with the lower convex envelope}
\label{vsm}
\end{figure}

Consider the case when $K = 100$ and $z = 5$. For every $m$ and $b$ such that $mb = 100$ Theorem~\ref{theorem1} shows a solution. Fig.~\ref{ratevsm} shows the plot of achievable rate $R$ against normalized cache memory size $M/N$ and Fig.~\ref{subvsm} shows the plot of base $10$ logarithm of subpacketization level against the $M/N$. In Fig.~\ref{ratevsm}, the point $(0,100)$ is trivially achieved, the point $(1/50,45)$ is achieved for $b = 50$, the point $(1/25,20)$ is achieved for $b = 25$, the point $(1/20,15)$ is achieved for $b = 20$, the point $(1/10,5)$ is achieved for $b = 10$, and the point $(1/5,0)$ is achieved for any legitimate value of $b$. To note that the user-to-cache association bipartite graph depends upon $b$. So, for instance, the user-to-cache association bipartite graph when $b = 50$ is different from the user-to-cache association bipartite graph when $b = 25$.

Fig.~\ref{ratevsm} shows that our scheme may achieve a better rate than SPE scheme, SR1 scheme, SR2 scheme, MR scheme, and RK scheme for some ranges of $M/N$. On the other hand, Figure~\ref{subvsm} shows that our scheme achieves a lesser subpacketization level than NT scheme, SICPS scheme, and RK scheme for some values of $M/N$.

At $M/N = 0.1$ our scheme achieves $R = 5$ files, NT scheme achieves $R \approx 4.5$ files, SICPS scheme achieves $R \approx 4.3$ files, while at the same value of $M/N$ the three schemes achieve a subpacketization level $10^{10}$, $10^{12.8}$, $10^{11.1}$ respectively (approximate values for the latter two).

While the memory-rate trade-off curve of the NT scheme and the SICPS scheme (In Fig.~\ref{ratevsm}) is below our scheme for the entire range of $M/N$, the RK scheme and the SR1 scheme seems to have a crossover point after $M/N = 0.16$. For clarity, we show these proximal points in Table~\ref{table2}.
\begin{table}[!htbp]\label{table2}
\caption{Some proximal points in Fig.~\ref{ratevsm} shown for clarity.}
\centering
%\begin{tabularx}{\textwidth}{|>{\centering$}p{1cm}<{$} | >{\centering$}p{2.67cm}<{$} | >{\raggedright\arraybackslash}X|}
\begin{tabularx}{0.41\textwidth}{ | >{\centering$}p{1cm}<{$}| >{\centering$}p{1.4cm}<{$} | >{\centering$}p{1.3cm}<{$} | >{\centering\arraybackslash$}X<{$} | }
\hline
\text{M/N} & \text{SR1 Scheme} & \text{RK Scheme} & \text{Our Scheme}\\
\hline
0.16 & 3.4965 & 4 & 2\\\hline
0.17 & 1.6953  & 2.25 & 1.5\\\hline
0.18 & 0.9528 & 1 & 1\\\hline
0.19 & 0.2103 & 0.25 & 0.5\\\hline
0.2 & 0 & 0 & 0\\\hline
\end{tabularx}
\end{table}

%In the plot shown in Figure~\ref{ratevsm} the memory rate trade-off curve of our scheme has $4$ non-trivial corner points, namely $(\frac{1}{50},45), (\frac{1}{25},20), (\frac{1}{20},15), (\frac{1}{10},5)$, and two trivial corner points, namely $(0,100), (\frac{1}{5},0)$. Since the user-to-cache association depends upon $b$, each of the four non-trivial corner points are achievable for different user-to-cache associations. For example, if $\frac{M}{N} = \frac{1}{50}$, then setting $b = 50, m = 2$ gives a set of user-to-cache associations (satisfying conditions \textbf{C1, C2, C3}), and for these associations we achieve a rate of $45$ files. If instead we had $\frac{M}{N} = \frac{1}{25}$, then setting $b = 25$ gives a different set of user-to-cache associations for which we achieve a rate of $20$ files. Note that the number of users and caches does not change with changing $
%\frac{M}{N}$ and remains fixed at $K$.

% value of $K$ does not change, so the natural resources remains the same.

%Now, if changing the user to cache association is not allowed while changing $\frac{M}{N}$, then this means that $b$ is fixed. In this case Theorem~2 of our paper provides a achievability scheme for all $\frac{M}{N}$ with $3$ non-trivial points, namely $(\frac{1}{b}, b -z)$, $(\frac{t}{b}, b - xz)$, $(\frac{b - (z-1)t}{b},0)$ where $t = \floor{\frac{b}{z}}$, which results in a memory rate trade-off curve with three linear segments. Furthermore, for a fixed $b$, the subpacketization level remains fixed at $b^m$ irrespective of $\frac{M}{N}$.

In Table~\ref{table3} we show the achievable rate and subpacketization level of all the
 relevant schemes. It has been shown in \cite{cheng} that their NT scheme performs at least as good as the scheme in \cite{hachem} in both achievable rate and subpacketization level for all memory sizes, and hence the latter scheme is not considered for comparison. Some of the equations in Table~\ref{table3} are sizable, in such cases the reader has been referred to see the equation in the original paper. 

%The scheme shown in \cite{anjana} is addressed separately.

For the rest of this section $t \in \{1,2,\ldots,\floor{\frac{b}{z}}\}$, $t^\prime \in \{1,2,\ldots,\floor{\frac{K}{z}}\}$, and $t^{\prime\prime} \in \{1,2,\ldots,K\}$. It is to be noted that in Table~\ref{table3} the SR2 scheme is applicable only if $t^{\prime\prime}$ divides $K$ and $(K - t^{\prime\prime}z + t^{\prime\prime})$ divides $K$; and SR1 scheme is applicable only if $gcd(t^{\prime\prime},K) = 1$.
\begin{table}[!htbp]\label{table3}
\caption{Comparison of schemes}
\centering
%\begin{tabularx}{\textwidth}{|>{\centering$}p{1cm}<{$} | >{\centering$}p{2.67cm}<{$} | >{\raggedright\arraybackslash}X|}
%\begin{tabularx}{\textwidth}{| >{\raggedright\arraybackslash}X|} | >{\centering$}p{1cm}<{$} | >{\centering$}p{1cm}<{$}| >{\centering$}p{1.3cm}<{$} | >{\centering$}p{1.3cm}<{$} | >{\centering\arraybackslash$}X<{$} | }
\begin{tabularx}{0.8\textwidth}{| >{\centering\arraybackslash}p{2cm} | >{\centering\arraybackslash$}p{2cm}<{$} | >{\centering\arraybackslash$}X<{$}|  >{\centering\arraybackslash$}X<{$}|}
\hline
 & \text{M/N} & \text{Rate} & \text{Subpacketization level}\\
 \hline
Our Scheme & \frac{t}{b} & b - tz & b^m\\\hline
SPE Scheme & \frac{2}{K} & \text{see eqn. (2) of \cite{serbetci}} & \frac{K(K-2z+2)}{4}\\\hline
RK Scheme & \frac{t^\prime}{K} & \frac{(K - t^\prime z)^2}{K} & \frac{K}{t^\prime}\binom{K - t^\prime z + t^\prime - 1}{t^\prime - 1}\\\hline
NT Scheme & \text{same as above} & \frac{K-t^\prime z}{t^\prime + 1} & K\binom{K - t^\prime z + t^\prime}{t^\prime}\\\hline
SICPS Scheme & \text{same as above} & \text{see eqn. (6) of \cite{reddy2}} & \frac{K}{t^\prime}\binom{K - t^\prime z + t^\prime - 1}{t^\prime - 1}\\\hline
SR1 Scheme & \frac{t^{\prime\prime}}{K} & \text{see Theorem~1 of \cite{shanuja}} & cK\, (1\leq c\leq K)\\
\hline
SR2 Scheme & \text{same as above} & \frac{(K - t^{\prime\prime} z)(K - t^{\prime\prime} z + t^{\prime\prime})}{2K} & K\\\hline
MR Scheme & \frac{1}{K} & \frac{1}{K}\ceil{\frac{K(K-z)}{2 + \floor{\frac{z}{K - z + 1}} + \floor{\frac{z-1}{K - z + 1}}}} & K\\\hline
\end{tabularx}
\end{table}

\subsubsection{Comparison with the SPE scheme}
Due to the complexity of equation (2) of \cite{serbetci} we could not provide an analytical comparison between the achievable rate of the SPE scheme and our scheme. For the case considered in Fig.~\ref{ratevsm} it can be seen that our scheme achieves a lesser rate for a large range of the normalized memory size. %, the SPE scheme achieves a lesser subpacketization level.

%For $\frac{M}{N} = \frac{2}{K}$ the SPE scheme achieves a coding gain between $3$ and $4$; whereas setting $\frac{t}{b} = \frac{2}{K}$ we find that for $t$ to be an integer $m$ (which is the coding gain of our scheme) can be either $1$ or $2$. 

%We now compare the subpacketization level of the SPE scheme and our scheme. If we take $m = 1$, then $b = K$ and at $\frac{M}{N} = \frac{2}{K}$ our scheme achieves a lesser (linear) subpacketization  level than the SPE scheme (though it is at the cost of a lesser coding gain and thus a higher rate). %If $m = 2$, then $b = \frac{K}{2}$ and at $\frac{M}{N} = \frac{2}{K}$ our scheme achieves a subpacketization level than the SPE scheme (though very likely it is at the cost of a higher rate). 

For a comparison of the subpacketization level, if in our scheme we take $m = 1$, then $b = K$ and at $\frac{M}{N} = \frac{2}{K}$ our scheme achieves a lesser (linear) subpacketization  level than the SPE scheme (though it is at the cost of a lesser coding gain and thus a higher rate).

%For the case considered in Fig.~\ref{ratevsm} and Fig.~\ref{subvsm} it can be seen that while our scheme achieves a lesser rate for a large range of the normalized memory size, the SPE scheme achieves a lesser subpacketization level.

\subsubsection{Comparison with the RK scheme, NT scheme, and SICPS scheme}
\textit{Comparison of the achievable rate:}
It has been shown in \cite{reddy2} that the achievable rate of the SICPS scheme is lesser than that of the RK scheme and NT scheme. However, due to the complexity of equation~(6) of \cite{reddy2} prescribing the achievable rate of the SICPS scheme, we are not able to provide an analytical comparison. For the example considered in Fig.~\ref{ratevsm}, it can be seen that the achievable rate of the NT scheme and the SICPS scheme is less than our scheme, but the rate of the RK scheme is higher than our scheme for the most part of the memory region. 
\begin{lemma}\label{P:RK}
At $\frac{M}{N} = \frac{t}{b} = \frac{t^\prime}{K}$ where $\frac{M}{N} < min\{\frac{1}{b}\floor{\frac{b}{z}}, \frac{1}{K}\floor{\frac{K}{z}}, \frac{K - b}{Kz}\}$ our scheme achieves a lesser rate than that of the RK scheme.
\end{lemma}
The proof is shown in Appendix~\ref{p:RK}.

At $\frac{M}{N} = \frac{t}{b} = \frac{t^\prime}{K}$ where $\frac{t}{b} < min\{\frac{1}{b}\floor{\frac{b}{z}}, \frac{1}{K}\floor{\frac{K}{z}}\}$, the NT scheme achieves a rate $\frac{m(b - tz)}{mt + 1}$, whereas our scheme achieves a rate $(b - tz)$. So the achievable rate of the NT scheme is lesser by a factor of $(t + \frac{1}{m}) = b(\frac{M}{N} + \frac{1}{K})$. So for a given $\frac{M}{N}$ and $K$, the less the value of $b$ the less is the gap between the rate of our scheme and the NT scheme. %However, it can be seen that reducing $b$ in our scheme increases $m$ and thereby the subpacketization level.

\textit{Comparison of the Subpacketization level:}
We show that even though the SICPS scheme and the NT scheme achieve a lesser rate than our scheme, at least for  some values $\frac{M}{N}$ our scheme achieves a lesser subpacketization level. Analytical comparison of the subpacketization level of these two schemes with our schemes is complicated due to the disparate nature of the closed form expression of the respective subpacketization levels, and hence we could provide an analytical result only under a restricted setting. 
\begin{lemma}\label{P:SICPS}
At $\frac{M}{N} = \frac{1}{b} = \frac{m}{K}$ if $b \geq \sqrt{K(z-1)} + 1$, $b > z$, and $m \leq \floor{\frac{K}{z}}$ our scheme achieves a lesser subpacketization level than the RK scheme, NT scheme, and the SICPS scheme.
\end{lemma}  
The proof is shown in Appendix~\ref{p:SICPS}. For the case considered in Fig.~\ref{subvsm} it can be seen that our scheme achieves a lesser subpacketization level than these two schemes at several values $\frac{M}{N}$.

\subsubsection{Comparison with the SR1 scheme}
For the SR1 scheme the non-trivial corner points has $\frac{M}{N} = \frac{t^{\prime\prime}}{K}$. For our scheme the non-trivial corner points are $\frac{M}{N} = \frac{t}{b} = \frac{mt}{K}$. So for the $\frac{M}{N}$ value of the two schemes to coincide we must have $\frac{t^{\prime\prime}}{K} = \frac{t}{b}$, which leads to $t^{\prime\prime} = mt$. However, in such a case $gcd(t^{\prime\prime},K) = gcd(mt,mb) \geq m$. So unless $m = 1$ and $gcd(t,b) = 1$, the $\frac{M}{N}$ value of the two schemes do not coincide. 

%For $b = K$, $m = 1$, our scheme achieves a subpacketization level of $K$, which is same or lesser than the subpacketization l

Due to the complexity of the achievable rate expression of the SR1 scheme and its disparity with our scheme, we could only provide the following analytical comparison under a restricted setting.
\begin{lemma}\label{lemmasr1}
For some positive integers $m_1, m_2$ such that both $m_1$ and $m_2$ divides $K$, $m_1 < m_2$, $b_1 = \frac{K}{m_1}$, $b_2 = \frac{K}{m_2}$, $b_1, b_2 \geq z$, $\lambda\in \mathbb{R}$, $0 \leq \lambda \leq 1$, $t^{\prime\prime} = m_1 + \lambda(m_2 - m_1)$, $t^{\prime\prime} \neq \frac{K - 1}{z}$, $gcd(t^{\prime\prime},K) = 1$,  at $\frac{M}{N} = \frac{t^{\prime\prime}}{K}$ our scheme achieves a better rate than the SR2 scheme if
\begin{IEEEeqnarray*}{l}
b_1 + \lambda(b_2 - b_1) \leq \frac{(K - t^{\prime\prime}z)(K - t^{\prime\prime}z + 2)}{K+2} + z.
\end{IEEEeqnarray*}
\end{lemma}
The proof is shown in Appendix~\ref{p:SR1}.

For example, for $K = 100$, $z = 5$, if we take $m_1 = 4$, $m_2 = 10$, $\lambda = \frac{1}{2}$, then all the conditions of Lemma~\ref{lemmasr1} gets satisfied, and for the point $\frac{t^{\prime\prime}}{K} = \frac{7}{100}$ the lemma claims that our scheme achieves a lesser rate. Indeed, at $\frac{M}{N} = \frac{7}{100}$, the SR1 scheme a rate of $32$ files, whereas our scheme achieves a rate of $12.5$ files (the point $(\frac{7}{100},12.5)$ lies on the line connecting the points $(\frac{1}{25},20)$ and $(\frac{1}{10},5)$). (The comparison is shown in Fig.~\ref{ratevsm}.)

\subsubsection{Comparison with the SR2 scheme}
\begin{lemma}\label{P:SR2}
At $\frac{M}{N} = \frac{t}{b} = \frac{t^{\prime\prime}}{K}$ where $t$ divides $b$, $(b - tz + t)$ divides $b$, and $t \leq \floor{\frac{b}{z}}$ our scheme achieves a lesser rate than that of the SR2 scheme when $\frac{M	}{N} \leq \frac{m - 2}{m(z-1)}$.
\end{lemma}
The proof is shown in Appendix~\ref{p:SR2}.

For example, for $K = 120$, $z = 5$, at $\frac{M}{N} = \frac{15}{120}$, the SR1 scheme achieves a rate $11.25$ files, whereas our scheme for $t = 3, b = 24, m = 5$ achieves a rate $9$ files. 
For the case considered in Fig.~\ref{ratevsm}, for no $\frac{M}{N} = \frac{t}{b}$ the conditions $t$ divides $b$ and $(b - tz + t)$ divides $b$ satisfied, so the plot of the rate achieved by the SR2 scheme is a straight line connecting the points $(0,100)$ and $(1/5,0)$.

The SR2 scheme has a subpacketization level of $K$. For $b = K$ and $m = 1$ our scheme also produces a scheme with subpacketization level $K$. If the subpacketization of our scheme and the SR2 scheme is kept the same, then SR2 scheme achieves a lesser rate.

\subsubsection{Comparison with the MR scheme}
\begin{lemma}\label{P:MR}
At $\frac{M}{N} = \frac{1}{b}$ where $K = mb$ and $m \geq 3$, our scheme achieves a lesser rate than the MR scheme.
\end{lemma}
The proof is shown in Appendix~\ref{p:MR}. For $m = 2$ the rates of our scheme and the MR scheme are equal. In Fig.~\ref{ratevsm} it can be seen that for $\frac{M}{N} > \frac{1}{50}$ our scheme achieves a lesser rate.

 %In Fig.~\ref{ratevsm} it can be seen that from $\frac{M}{N} = \frac{1}{50}$ onwards our scheme achieves a lesser rate.

%\subsection{Disadvantages of our scheme}
%%Our scheme has two disadvantages. 
%First, our scheme achieves a rate of $0$ files at $\frac{M}{N} = \frac{b - (z-1)t}{b}$ where $t = \floor{\frac{b}{z}}$. Whereas with coded placement a rate of $0$ files can be achieved at $\frac{M}{N} = \frac{1}{z}$, and with uncoded placement the same can be achieved at $\frac{M}{N} = \frac{1}{K}\ceil{\frac{K}{z}}$. For example, for $K = 100$, $b = 20$, $z = 7$, the rate of our scheme
%is non-zero for $\frac{M}{N} < \frac{8}{20}$, whereas, with uncoded placement the rate goes to zero at $\frac{M}{N} = \frac{3}{20}$. 
%%
%
%Second, if one sets the condition that $z \geq \frac{K}{n}$, then as $b \geq z$, we have $b \geq \frac{mb}{n}$, which leads to $m \leq n$. So the coding gain of our scheme must be less than or equal to $n$ in such case.
%
%%\end{enumerate}
%%new line here
%Third, the number of corner points depends on how many factors $K$ has. However, for example, if $K$ is prime and greater than $9$, then $K$ can be written as a sum of two non-prime integers $9$ and $K-9$ (where $K-9$ is even). Then, we can solve the problem by treating it as two separate MACC problems (one with $9$ users and caches, the other with $K-9$ users and caches).

\section{Examples}\label{example}
%\begin{example}\label{ex-fm}
\subsection{First Example: \texorpdfstring{$K = 8$, $z = 2$, $\frac{M}{N} = \frac{1}{4}$}{K = 8, z = 2, M/N = 1/4}}
In the first example, we consider a MACC problem with $K = 8$, $z = 2$, $\frac{M}{N} = \frac{1}{4}$. For $b = 4$, $m = 2$, let $G = (U,C,E)$ shown in Fig.~\ref{graph1} be the user-to-cache association bipartite graph. It is to be noted that there are other user-to-cache association bipartite graphs as well that satisfies the criteria set by this example problem. We have $U = K_1 \cup K_2$ where $K_1 = \{k(1,1), k(1,2), k(1,3), k(1,4)\}$, $K_2 = \{k(2,1), k(2,2), k(2,3), k(2,4)\}$. And, $C = C_1 \cup C_2$ where $C_1 = \{c(1,1), c(1,2), c(1,3), c(1,4)\}$, $C_2 = \{c(2,1), c(2,2), c(2,3), c(2,4)\}$.

%Consider the MACC problem whose user-to-cache association is given by the bipartite graph shown in Fig.~\ref{graph1}. This problem has $8$ users, $8$ caches, $z = 2$, and the user-to-cache association bipartite graph satisfies conditions \textbf{C1, C2, C3} for $b = 4$, $m = 2$. 

Condition \textbf{C3} ensures the existence of a bijection $f_{M_i}$ from $K_i$ to $C_i$. Let us consider the following bijection $f_{M_1}(k(1,1)) = c(1,1)$, $f_{M_1}(k(1,2)) = c(1,2)$, $f_{M_1}(k(1,3)) = c(1,4)$, $f_{M_1}(k(1,4)) = c(1,3)$, $f_{M_2}(k(2,1)) = c(2,4)$, $f_{M_2}(k(2,2)) = c(2,3)$, $f_{M_2}(k(2,3)) = c(2,2)$, $f_{M_2}(k(2,4)) = c(2,1)$. 

The map $f_{M_i}$ could have been defined in other ways as well. For instance, we could have defined $f_{M_1}(k(1,1)) = c(1,3)$, $f_{M_1}(k(1,2)) = c(1,4)$, $f_{M_1}(k(1,3)) = c(1,1)$, $f_{M_1}(k(1,4)) = c(1,2)$.
%\end{example}

As per condition \textbf{C2}, $C_i = C_{(i,1)} \cup C_{(i,2)}$, where
\begin{IEEEeqnarray*}{l}
C_{(i,1)}  = \{c(i,1), c(i,2)\}, C_{(i,2)} = \{c(i,3), c(i,4)\} \text{ for } 1\leq i\leq 2.
\end{IEEEeqnarray*}
For $i = 1,2$, $j=1,2,3,4$, set $C_{k(i,j)}$ denotes the set of caches user $k(i,j)$ accesses. We have
\begin{IEEEeqnarray*}{l}
C_{k(1,1)} = \{c(1,1), c(1,3)\}, C_{k(1,2)} = \{c(1,2), c(1,4)\},  
C_{k(1,3)} = \{c(1,1), c(1,4)\}, C_{k(1,4)} = \{c(1,2), c(1,3)\},\\ 
C_{k(2,1)} = \{c(2,1), c(2,4)\}, C_{k(2,2)} = \{c(2,2), c(2,3)\},   
C_{k(2,3)} = \{c(2,2), c(2,4)\}, C_{k(2,4)} = \{c(2,1), c(2,3)\}. 
\end{IEEEeqnarray*}
%\begin{example}\label{ex-2}
%For the user-to-cache association bipartite graph shown in Fig.~\ref{graph1}, 

%Since $m = 2$ and $b = 4$, following the construction procedure shown in the proof of Theorem~\ref{theorem-1}, 
We consider the following MCRD to be used for placement and delivery.
%
%We consider the following MCRD for placement. % (MCRD property proved in Example~\ref{ex-a3}).% the corresponding MCRD
\begin{IEEEeqnarray*}{l}
\mathcal{P}_1 = \{ \{1,2,3,4\}, \{5,6,7,8\}, \{9,10,11,12\}, \{13,14,15,16\} \}\\
\mathcal{P}_2 = \{ \{1,5,9,13\}, \{2,6,10,14\}, \{3,7,11,15\}, \{4,8,12,16\}\}.
\end{IEEEeqnarray*}
It is to be noted that we could have used any MCRD which has $m = 2$ parallel classes, $b = 4$ blocks in each parallel class, and $\mu_2 = 1$. Let
\begin{IEEEeqnarray*}{l}
B(1,1) = \{1,2,3,4\},  B(1,2) = \{5,6,7,8\},  B(1,3) = \{9,10,11,12\}, B(1,4) = \{13,14,15,16\}\\
B(2,1) = \{1,5,9,13\}, B(2,2) = \{2,6,10,14\}, B(2,3) = \{3,7,11,15\},  B(2,4) = \{4,8,12,16\}.
\end{IEEEeqnarray*}
%
%\end{example}
%\begin{example}\label{ex-3}
%For the user-to-cache association bipartite graph shown in Fig.~\ref{graph1} 
%
%
Partition sets $\mathcal{P}_1$ and $\mathcal{P}_2$ as following 
\begin{IEEEeqnarray*}{l}
\mathcal{P}_{(i,1)} = \{B(i,1), B(i,2)\}, \mathcal{P}_{(i,2)} = \{B(i,3), B(i,4)\} \text{ for } 1\leq i\leq 2.
\end{IEEEeqnarray*}
%We consider the cases when $t \leq b - (z-1)\floor{b/z} = 2$.
%
%\newline\textbf{Case II.1:} \textbf{$t = 1$}. We have $t^\prime = t_z = 1$.
%Let $\frac{M}{N} = \frac{t}{b} = \frac{t}{4}$ where $t \in \mathbb{N}$. 

% The definition of $B_{c(i,j)}$ depends upon the value of $\frac{M}{N}$. %In the following, we following we consider all values of $t$ and show the corresponding placement (decided by $B_{c(i,j)}$) and delivery.
%
%\begin{case}
%$t = 1$.
%\end{case}
\textit{Placement:} Split each file into $b^m = 16$ subfiles. A cache stores the subfiles indexed by the elements of some of the blocks of the MCRD considered above. Let the cache $c(i,j)$ store the subfiles (of all files) indexed by the elements of the blocks contained in $B_{c(i,j)}$, where
\begin{IEEEeqnarray}{l}
B_{c(i,j)} = \{B(i,j)\} \text{ for } 1\leq i\leq 2, 1\leq j\leq 4.\label{case2}
\end{IEEEeqnarray}
%
%\end{example} 
%
%\begin{example}\label{ex-4}
%For example~\ref{ex-3} 
%We have the following for $t=1$.
Let $B_{k(i,j)}$ denote the set of blocks such that if $B \in B_{k(i,j)}$ then for some cache $c(i,j^\prime) \in C_{k(i,j)}$ we have $B \in B_{c(i,j^\prime)}$. So we have
%whose contents are the indices of the subfiles stored in the caches accessed by user $k(i,j)$.
\begin{IEEEeqnarray*}{llll}
B_{k(1,1)} = \{B(1,1), B(1,3)\},\qquad &B_{k(1,2)} = \{B(1,2), B(1,4)\}, \qquad
&B_{k(1,3)} = \{B(1,1), B(1,4)\},\\ B_{k(1,4)} = \{B(1,2), B(1,3)\},
&B_{k(2,1)} = \{B(2,1), B(2,4)\}, &B_{k(2,2)} = \{B(2,2), B(2,3)\},\\
B_{k(2,3)} = \{B(2,2), B(2,4)\}, &B_{k(2,4)} = \{B(2,1), B(2,3)\}.
% \text{ for } 1\leq i\leq 2.
\end{IEEEeqnarray*}
%
%\end{example}
%
\begin{figure}[h]
\centering
\includegraphics[width=0.9\textwidth]{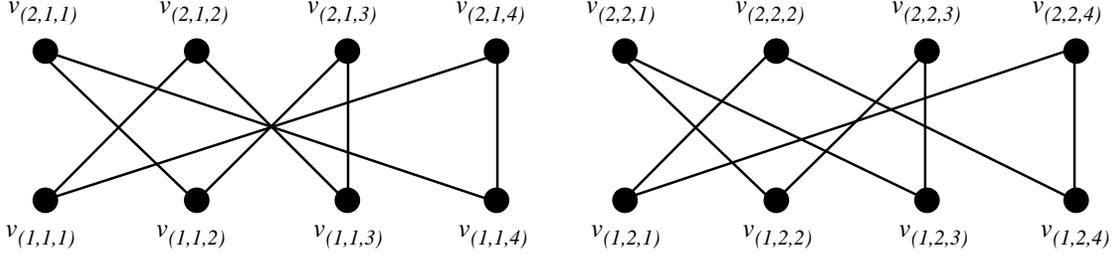}
\caption{The demand graph $\bar{G} = (V_1,V_2,\bar{E})$ for the user-to-cache association bipartite graph $G = (U,C,E)$ shown in Fig.~\ref{graph1} when $\frac{M}{N} = \frac{1}{4}$ (set $B_{c(i,j)}$ defined in equation~(\ref{case2})). An edge connects $v_{(1,i,j)}$ to $v_{(2,i^\prime,j^*)}$ if and only if $i = i^\prime$ and the set $B_{f_{M_i}^{-1}(c(i,j))}$ does not contain the block $B(i,j^*)$.}
\label{graph2}
\end{figure}

%\begin{example}\label{ex-5}

We now construct a new graph that we call as the demand graph $\bar{G} = (V_1,V_2,\bar{E})$ shown in Fig.~\ref{graph2}. We have for $l = 1,2$, $V_l = \{v_{(l,i,j)} | 1\leq i\leq 2, 1\leq j\leq 4\}$. There is a bijection from the vertices in $V_1$ to the caches in $C$, and there is another bijection from the vertices in $V_2$ to the blocks in $\mathcal{P}_1 \cup \mathcal{P}_2$. Say for $i=1,2$, $1\leq j\leq 4$, the vertex $v_{(1,i,j)}$ maps to the cache $c(i,j)$, similarly, say that the vertex $v_{(2,i,j)}$ maps to the block $B(i,j)$. For $i=1,2$, $1\leq j\leq 4$ there exists an edge in $\bar{E}$ connecting vertex $v_{(1,i,j)} \in V_1$ to vertex $v_{(2,\bar{i},j^*)} \in V_2$ if and only if $i = \bar{i}$ and $B(i,j^*) \notin B_{k(i,j^\prime)}$ where $k(i,j^\prime) = f_{M_i}^{-1}(c(i,j))$.

We define a map $f_{(i,j)}: \{1,2\} \rightarrow \{1,2,3,4\}\setminus\{l \,|\, B(i,l) \in B_{k(i,j)}\} $ for $1\leq i\leq 2$, $1\leq j\leq 4$ as following.
\begin{IEEEeqnarray*}{llllll}
f_{(1,1)}(1) = 2,\quad & f_{(1,1)}(2) = 4,\quad & f_{(1,2)}(1) = 1,\quad & f_{(1,2)}(2) = 3,\quad
& f_{(1,3)}(1) = 2,\quad & f_{(1,3)}(2) = 3,\\f_{(1,4)}(1) = 1,& f_{(1,4)}(2) = 4,
&f_{(2,1)}(1) = 2,& f_{(2,1)}(2) = 3,& f_{(2,2)}(1) = 1,& f_{(2,2)}(2) = 4,\\
f_{(2,3)}(1) = 1,& f_{(2,3)}(2) = 3,& f_{(2,4)}(1) = 2,& f_{(2,4)}(2) = 4.
\end{IEEEeqnarray*}

We define a partial matching $M^n_{j_1,j_2}$ for $n = 1,2$, $1\leq j_1,j_2\leq 4$ as following.
\begin{IEEEeqnarray*}{l}
M^n_{j_1,j_2} = \{(v_{(1,i,j_i)}, v_{(2,i,f_{(i,j_i^\prime)}(n))})\, | \, i=1,2 \} \text{ where }
k(i,j_i^\prime) = f_{M_i}^{-1}(c(i,j_i)).
\end{IEEEeqnarray*}
So we have
%Consider the partial matching $M^1_{j_1,j_2}$ and $M^2_{j_1,j_2}$ for $1\leq j_1,j_2 \leq 4$ shown below (total $32$ matchings). %Note that we use $f_{M_i}$ as shown in Example~\ref{ex-fm}.
\begin{IEEEeqnarray*}{ll}
M^1_{1,1} = \{ (v_{(1,1,1)},v_{(2,1,2)}), (v_{(1,2,1)},v_{(2,2,2)})\}\qquad &
M^1_{1,2} = \{ (v_{(1,1,1)},v_{(2,1,2)}), (v_{(1,2,2)},v_{(2,2,1)})\}\\
M^1_{1,3} = \{ (v_{(1,1,1)},v_{(2,1,2)}), (v_{(1,2,3)},v_{(2,2,1)})\}&
M^1_{1,4} = \{ (v_{(1,1,1)},v_{(2,1,2)}), (v_{(1,2,4)},v_{(2,2,2)})\}\\
M^1_{2,1} = \{ (v_{(1,1,2)},v_{(2,1,1)}), (v_{(1,2,1)},v_{(2,2,2)})\}&
M^1_{2,2} = \{ (v_{(1,1,2)},v_{(2,1,1)}), (v_{(1,2,2)},v_{(2,2,1)})\}\\
M^1_{2,3} = \{ (v_{(1,1,2)},v_{(2,1,1)}), (v_{(1,2,3)},v_{(2,2,1)})\}&
M^1_{2,4} = \{ (v_{(1,1,2)},v_{(2,1,1)}), (v_{(1,2,4)},v_{(2,2,2)})\}\\
M^1_{3,1} = \{ (v_{(1,1,3)},v_{(2,1,1)}), (v_{(1,2,1)},v_{(2,2,2)})\}&
M^1_{3,2} = \{ (v_{(1,1,3)},v_{(2,1,1)}), (v_{(1,2,2)},v_{(2,2,1)})\}\\
M^1_{3,3} = \{ (v_{(1,1,3)},v_{(2,1,1)}), (v_{(1,2,3)},v_{(2,2,1)})\}&
M^1_{3,4} = \{ (v_{(1,1,3)},v_{(2,1,1)}), (v_{(1,2,4)},v_{(2,2,2)})\}\\
M^1_{4,1} = \{ (v_{(1,1,4)},v_{(2,1,2)}), (v_{(1,2,1)},v_{(2,2,2)})\}&
M^1_{4,2} = \{ (v_{(1,1,4)},v_{(2,1,2)}), (v_{(1,2,2)},v_{(2,2,1)})\}\\
M^1_{4,3} = \{ (v_{(1,1,4)},v_{(2,1,2)}), (v_{(1,2,3)},v_{(2,2,1)})\}&
M^1_{4,4} = \{ (v_{(1,1,4)},v_{(2,1,2)}), (v_{(1,2,4)},v_{(2,2,2)})\}\\
M^2_{1,1} = \{ (v_{(1,1,1)},v_{(2,1,4)}), (v_{(1,2,1)},v_{(2,2,4)})\}&
M^2_{1,2} = \{ (v_{(1,1,1)},v_{(2,1,4)}), (v_{(1,2,2)},v_{(2,2,3)})\}\\
M^2_{1,3} = \{ (v_{(1,1,1)},v_{(2,1,4)}), (v_{(1,2,3)},v_{(2,2,4)})\}&
M^2_{1,4} = \{ (v_{(1,1,1)},v_{(2,1,4)}), (v_{(1,2,4)},v_{(2,2,3)})\}\\
M^2_{2,1} = \{ (v_{(1,1,2)},v_{(2,1,3)}), (v_{(1,2,1)},v_{(2,2,4)})\}&
M^2_{2,2} = \{ (v_{(1,1,2)},v_{(2,1,3)}), (v_{(1,2,2)},v_{(2,2,3)})\}\\
M^2_{2,3} = \{ (v_{(1,1,2)},v_{(2,1,3)}), (v_{(1,2,3)},v_{(2,2,4)})\}&
M^2_{2,4} = \{ (v_{(1,1,2)},v_{(2,1,3)}), (v_{(1,2,4)},v_{(2,2,3)})\}\\
M^2_{3,1} = \{ (v_{(1,1,3)},v_{(2,1,4)}), (v_{(1,2,1)},v_{(2,2,4)})\}&
M^2_{3,2} = \{ (v_{(1,1,3)},v_{(2,1,4)}), (v_{(1,2,2)},v_{(2,2,3)})\}\\
M^2_{3,3} = \{ (v_{(1,1,3)},v_{(2,1,4)}), (v_{(1,2,3)},v_{(2,2,4)})\}&
M^2_{3,4} = \{ (v_{(1,1,3)},v_{(2,1,4)}), (v_{(1,2,4)},v_{(2,2,3)})\}\\
M^2_{4,1} = \{ (v_{(1,1,4)},v_{(2,1,3)}), (v_{(1,2,1)},v_{(2,2,4)})\}&
M^2_{4,2} = \{ (v_{(1,1,4)},v_{(2,1,3)}), (v_{(1,2,2)},v_{(2,2,3)})\}\\
M^2_{4,3} = \{ (v_{(1,1,4)},v_{(2,1,3)}), (v_{(1,2,3)},v_{(2,2,4)})\}&
M^2_{4,4} = \{ (v_{(1,1,4)},v_{(2,1,3)}), (v_{(1,2,4)},v_{(2,2,3)})\}.
\end{IEEEeqnarray*}
For each matchings $M^n_{j_1,j_2}$ for $n = 1,2$, $1\leq j_1,j_2 \leq 4$, we transmit one subfile $Y^n_{j_1,j_2}$. For a matching $M^n_{j_1,j_2} = \{(v_{(1,1,j_1)},v_{(2,1,j_1^*)}), (v_{(1,2,j_2)},v_{(2,2,j_2^*)})\}$, define $S^{n,1}_{j_1,j_2} = B(1,j_1^*) \cap B(2,j_2)$ and $S^{n,2}_{j_1,j_2} = B(1,j_1) \cap B(2,j_2^*)$. We transmit the following subfiles
\begin{IEEEeqnarray*}{l}
Y^n_{j_1,j_2} = W^{d_{f_{M_1}^{-1}(c(1,j_1))}}(S^{n,1}_{j_1,j_2}) + W^{d_{f_{M_2}^{-1}(c(2,j_2))}}(S^{n,2}_{j_1,j_2})\; \text{ for } 1\leq n\leq 2, 1\leq j_1,j_2 \leq 4.
\end{IEEEeqnarray*}
The transmissions $Y^n_{j_1,j_2}$ along with the set $S^{n,i}_{j_1,j_2}$ for $n,i = 1, 2$, $1\leq j_1,j_2 \leq 4$, are shown in the following.
\begin{IEEEeqnarray*}{lll}
S^{1,1}_{1,1} = B(1,2) \cap B(2,1) = \{5\}, \quad &S^{1,2}_{1,1} = B(1,1) \cap B(2,2) = \{2\},\qquad
&Y^1_{1,1} = W^{d_{(1,1)}}(5) + W^{d_{(2,4)}}(2)\IEEEyesnumber\label{jan1}\\
S^{1,1}_{1,2} = B(1,2) \cap B(2,2) = \{6\}, &S^{1,2}_{1,2} = B(1,1) \cap B(2,1) = \{1\},
&Y^1_{1,2} = W^{d_{(1,1)}}(6) + W^{d_{(2,3)}}(1)\\
S^{1,1}_{1,3} = B(1,2) \cap B(2,3) = \{7\}, &S^{1,2}_{1,3} = B(1,1) \cap B(2,1) = \{1\},
&Y^1_{1,3} = W^{d_{(1,1)}}(7) + W^{d_{(2,2)}}(1)\\
S^{1,1}_{1,4} = B(1,2) \cap B(2,4) = \{8\}, &S^{1,2}_{1,4} = B(1,1) \cap B(2,2) = \{2\},
&Y^1_{1,4} = W^{d_{(1,1)}}(8) + W^{d_{(2,1)}}(2)\\
S^{1,1}_{2,1} = B(1,1) \cap B(2,1) = \{1\}, &S^{1,2}_{2,1} = B(1,2) \cap B(2,2) = \{6\},
&Y^1_{2,1} = W^{d_{(1,2)}}(1) + W^{d_{(2,4)}}(6)\\
S^{1,1}_{2,2} = B(1,1) \cap B(2,2) = \{2\}, &S^{1,2}_{2,2} = B(1,2) \cap B(2,1) = \{5\},
&Y^1_{2,2} = W^{d_{(1,2)}}(2) + W^{d_{(2,3)}}(1)\\
S^{1,1}_{2,3} = B(1,1) \cap B(2,3) = \{3\}, &S^{1,2}_{2,3} = B(1,2) \cap B(2,1) = \{5\},
&Y^1_{2,3} = W^{d_{(1,2)}}(3) + W^{d_{(2,2)}}(5)\\
S^{1,1}_{2,4} = B(1,1) \cap B(2,4) = \{4\}, &S^{1,2}_{2,4} = B(1,2) \cap B(2,2) = \{6\},
&Y^1_{2,4} = W^{d_{(1,2)}}(4) + W^{d_{(2,1)}}(6)\\
S^{1,1}_{3,1} = B(1,1) \cap B(2,1) = \{1\}, &S^{1,2}_{3,1} = B(1,3) \cap B(2,2) = \{10\},
&Y^1_{3,1} = W^{d_{(1,4)}}(1) + W^{d_{(2,4)}}(10)\\
S^{1,1}_{3,2} = B(1,1) \cap B(2,2) = \{2\}, &S^{1,2}_{3,2} = B(1,3) \cap B(2,1) = \{9\},
&Y^1_{3,2} = W^{d_{(1,4)}}(2) + W^{d_{(2,3)}}(9)\\
S^{1,1}_{3,3} = B(1,1) \cap B(2,3) = \{3\}, &S^{1,2}_{3,3} = B(1,3) \cap B(2,1) = \{9\},
&Y^1_{3,3} = W^{d_{(1,4)}}(3) + W^{d_{(2,2)}}(9)\\
S^{1,1}_{3,4} = B(1,1) \cap B(2,4) = \{4\}, &S^{1,2}_{3,4} = B(1,3) \cap B(2,2) = \{10\},
&Y^1_{3,4} = W^{d_{(1,4)}}(4) + W^{d_{(2,1)}}(10)\\
S^{1,1}_{4,1} = B(1,2) \cap B(2,1) = \{5\}, &S^{1,2}_{4,1} = B(1,4) \cap B(2,2) = \{14\},
&Y^1_{4,1} = W^{d_{(1,3)}}(5) + W^{d_{(2,4)}}(14)\\
S^{1,1}_{4,2} = B(1,2) \cap B(2,2) = \{6\}, &S^{1,2}_{4,2} = B(1,4) \cap B(2,1) = \{13\},
&Y^1_{4,2} = W^{d_{(1,3)}}(6) + W^{d_{(2,3)}}(13)\\
S^{1,1}_{4,3} = B(1,2) \cap B(2,3) = \{7\}, &S^{1,2}_{4,3} = B(1,4) \cap B(2,1) = \{13\},
&Y^1_{4,3} = W^{d_{(1,3)}}(7) + W^{d_{(2,2)}}(13)\\
S^{1,1}_{4,4} = B(1,2) \cap B(2,4) = \{8\}, &S^{1,2}_{4,4} = B(1,4) \cap B(2,2) = \{14\},
&Y^1_{4,4} = W^{d_{(1,3)}}(8) + W^{d_{(2,1)}}(14)\\
S^{2,1}_{1,1} = B(1,4) \cap B(2,1) = \{13\}, &S^{2,2}_{1,1} = B(1,1) \cap B(2,4) = \{4\},
&Y^2_{1,1} = W^{d_{(1,1)}}(13) + W^{d_{(2,4)}}(4)\\
S^{2,1}_{1,2} = B(1,4) \cap B(2,2) = \{14\}, &S^{2,2}_{1,2} = B(1,1) \cap B(2,3) = \{3\},
&Y^2_{1,2} = W^{d_{(1,1)}}(14) + W^{d_{(2,3)}}(3)\\
S^{2,1}_{1,3} = B(1,4) \cap B(2,3) = \{15\}, &S^{2,2}_{1,3} = B(1,1) \cap B(2,4) = \{4\},
&Y^2_{1,3} = W^{d_{(1,1)}}(15) + W^{d_{(2,2)}}(4)\\
S^{2,1}_{1,4} = B(1,4) \cap B(2,4) = \{16\}, &S^{2,2}_{1,4} = B(1,1) \cap B(2,3) = \{3\},
&Y^2_{1,4} = W^{d_{(1,1)}}(16) + W^{d_{(2,1)}}(3)\\
S^{2,1}_{2,1} = B(1,3) \cap B(2,1) = \{9\}, &S^{2,2}_{2,1} = B(1,2) \cap B(2,4) = \{8\},
&Y^2_{2,1} = W^{d_{(1,2)}}(9) + W^{d_{(2,4)}}(8)\\
S^{2,1}_{2,2} = B(1,3) \cap B(2,2) = \{10\}, &S^{2,2}_{2,2} = B(1,2) \cap B(2,3) = \{7\},
&Y^2_{2,2} = W^{d_{(1,2)}}(10) + W^{d_{(2,3)}}(7)\\
S^{2,1}_{2,3} = B(1,3) \cap B(2,3) = \{11\}, &S^{2,2}_{2,3} = B(1,2) \cap B(2,4) = \{8\},
&Y^2_{2,3} = W^{d_{(1,2)}}(11) + W^{d_{(2,2)}}(8)\\
S^{2,1}_{2,4} = B(1,3) \cap B(2,4) = \{12\}, &S^{2,2}_{2,4} = B(1,2) \cap B(2,3) = \{7\},
&Y^2_{2,4} = W^{d_{(1,2)}}(12) + W^{d_{(2,1)}}(7)\\
S^{2,1}_{3,1} = B(1,4) \cap B(2,1) = \{13\}, &S^{2,2}_{3,1} = B(1,3) \cap B(2,4) = \{12\},
&Y^2_{3,1} = W^{d_{(1,4)}}(13) + W^{d_{(2,4)}}(12)\\
S^{2,1}_{3,2} = B(1,4) \cap B(2,2) = \{14\}, &S^{2,2}_{3,2} = B(1,3) \cap B(2,3) = \{11\},
&Y^2_{3,2} = W^{d_{(1,4)}}(14) + W^{d_{(2,3)}}(11)\\
S^{2,1}_{3,3} = B(1,4) \cap B(2,3) = \{15\}, &S^{2,2}_{3,3} = B(1,3) \cap B(2,4) = \{12\},
&Y^2_{3,3} = W^{d_{(1,4)}}(15) + W^{d_{(2,2)}}(12)\\
S^{2,1}_{3,4} = B(1,4) \cap B(2,4) = \{16\}, &S^{2,2}_{3,4} = B(1,3) \cap B(2,3) = \{11\},
&Y^2_{3,4} = W^{d_{(1,4)}}(16) + W^{d_{(2,1)}}(11)\\
S^{2,1}_{4,1} = B(1,3) \cap B(2,1) = \{9\}, &S^{2,2}_{4,1} = B(1,4) \cap B(2,4) = \{16\},
&Y^2_{4,1} = W^{d_{(1,3)}}(9) + W^{d_{(2,4)}}(16)\\
S^{2,1}_{4,2} = B(1,3) \cap B(2,2) = \{10\}, &S^{2,2}_{4,2} = B(1,4) \cap B(2,3) = \{15\},
&Y^2_{4,2} = W^{d_{(1,3)}}(10) + W^{d_{(2,3)}}(15)\\
S^{2,1}_{4,3} = B(1,3) \cap B(2,3) = \{11\}, &S^{2,2}_{4,3} = B(1,4) \cap B(2,4) = \{16\},
&Y^2_{4,3} = W^{d_{(1,3)}}(11) + W^{d_{(2,2)}}(16)\\
S^{2,1}_{4,4} = B(1,3) \cap B(2,4) = \{12\}, &S^{2,2}_{4,4} = B(1,4) \cap B(2,3) = \{15\},
&Y^2_{4,4} = W^{d_{(1,3)}}(12) + W^{d_{(2,1)}}(15).
\end{IEEEeqnarray*}
Each transmission benefits two users, for instance, from equation~(\ref{jan1}), since $2 \in B(1,1)$, $B(1,1) \in B_{c(1,1)}$, $c(1,1) \in C_{k(1,1)}$, user $k(1,1)$ knows $W^{d_{(2,4)}}(2)$, so it can retrieve $W^{d_{(1,1)}}(5)$; similarly, and since $5 \in B(2,1)$, $B(2,1) \in B_{c(2,1)}$, $c(2,1) \in C_{k(2,4)}$, user $k(2,4)$ knows $W^{d_{(1,1)}}(5)$, so it can retrieve $W^{d_{(2,4)}}(2)$. After all the above transmissions, it can be verified that all the users can retrieve its demanded files.
%
%\end{example}
%\begin{example}
%In Example~\ref{ex-3}, 
%\begin{case}
%$t = 2$.
%\end{case}
%\textit{Placement:} Place contents as per the following.
%\begin{IEEEeqnarray*}{l}
%B_{c(i,1)} = B_{c(i,2)} = \{B(i,1), B(i,2)\} \text{ for } 1\leq i\leq 2\\
%B_{c(i,3)} = B_{c(i,4)} = \{B(i,3), B(i,4)\} \text{ for } 1\leq i\leq 2.
%\end{IEEEeqnarray*}
%So we have
%\begin{IEEEeqnarray*}{l}
%B_{k(i,j)} = \{B(i,1), B(i,2), B(i,3), B(i,4)\} \text{ for } 1\leq i\leq 2, 1\leq j\leq 4. 
%\end{IEEEeqnarray*}
%%
%In this case, it can be seen that any user can retrieve any file it demands from the contents of the cache memories it accesses.

%\end{example}
%\setcounter{case}{0}

%\subsection{Second Example: MACC problem with $K = 10$}

\subsection{Second Example: \texorpdfstring{$K = 14$, $z = 3$, $\frac{M}{N} = \frac{2}{7}$}{K = 14, z = 3, M/N = 2/7}}\label{sec:ex:2}
\begin{figure}[h]
\centering
\includegraphics[width=0.7\textwidth]{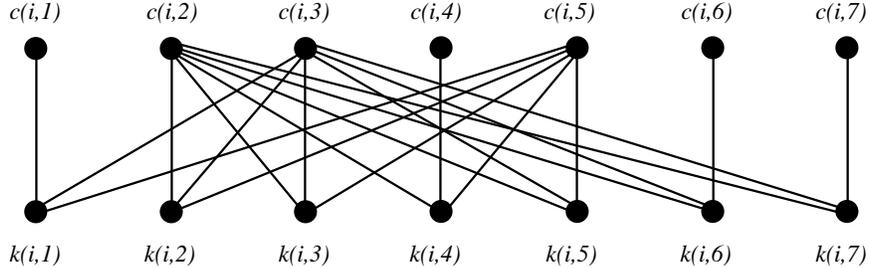}
\caption{Let the graph shown in the figure be denoted by $J_i$. The user-to-cache association bipartite graph $J = (U,C,E)$ considered in the second example in Section~\ref{sec:ex:2} is the disjoint union of $J_1$ and $J_2$. It can be verified that $J = (U,C,E)$ satisfies conditions \textbf{C1, C2, C3} for $b = 7$, $m = 2$, $z = 3$.}
\label{graphj}
\end{figure}
In the second example, we consider a MACC problem with $K = 14$, $z = 3$, $\frac{M}{N} = \frac{2}{7}$. For $b = 7$, $m = 2$, $z = 3$, let the user-to-cache association bipartite graph be $J = (U,C,E)$, an abstraction of which is shown in Fig.~\ref{graphj}. It is to be noted that there are other user-to-cache association bipartite graphs that satisfies the criteria set by this example problem. We have $U = K_1 \cup K_2$ where for $i = 1,2$, $K_i = \{k(i,j)\,| \, 1\leq j\leq 7\}$. And, $C = C_1 \cup C_2$ where for $i=1,2$, $C_i = \{c(i,j) | 1\leq j \leq 7\}$.

Let us consider the following definition of $f_{M_i}$: for $i=1,2$, $1\leq j\leq 7$, $f_{M_i}(k(i,j)) = c(i,j)$. As per condition \textbf{C2}, $C_i = C_{(i,1)} \cup C_{(i,2)} \cup C_{(i,3)}$, where
\begin{IEEEeqnarray*}{l}
C_{(i,1)}  = \{c(i,1), c(i,2)\}, C_{(i,2)} = \{c(i,3), c(i,4)\}, C_{(i,3)} = \{c(i,5), c(i,6), c(i,7)\} \text{ for } 1\leq i\leq 2.
\end{IEEEeqnarray*}
For $i = 1,2$, $1\leq j\leq 7$, set $C_{k(i,j)}$ denotes the set of caches user $k(i,j)$ accesses. We have
\begin{IEEEeqnarray*}{lll}
C_{k(i,1)} = \{c(i,1), c(i,3), c(i,5)\},\quad & C_{k(i,2)} = \{c(i,2), c(i,3), c(i,5)\},\quad
& C_{k(i,3)} = \{c(i,2), c(i,3), c(i,5)\},\\
C_{k(i,4)} = \{c(i,2), c(i,4), c(i,5)\},
& C_{k(i,5)} = \{c(i,2), c(i,3), c(i,5)\}, &C_{k(i,6)} = \{c(i,2), c(i,3), c(i,6)\},\\
C_{k(i,7)} = \{c(i,2), c(i,3), c(i,7)\}.
\end{IEEEeqnarray*}
%\begin{example}\label{ex-2}
%For the user-to-cache association bipartite graph shown in Fig.~\ref{graph1}, 

%Since $m = 2$ and $b = 4$, following the construction procedure shown in the proof of Theorem~\ref{theorem-1}, 
We consider the MCRD with the following parallel classes to be used for placement and delivery.
%
%We consider the following MCRD for placement. % (MCRD property proved in Example~\ref{ex-a3}).% the corresponding MCRD
%\begin{IEEEeqnarray*}{l}
%\mathcal{P}_i = \{B(i,j)\, |\,  1\leq j\leq 7  \} \text{ for } i = 1,2 \text{ where}
%\end{IEEEeqnarray*}
\begin{IEEEeqnarray*}{lll}
\IEEEeqnarraymulticol{3}{l}{\mathcal{P}_i = \{B(i,j)\, |\,  1\leq j\leq 7  \} \text{ for } i = 1,2 \text{ where}}\\*
B(1,1) = \{1,2,3,4,5,6,7\}, &B(1,2) = \{8,9,10,11,12,13,14\}, &B(1,3) = \{15,16,17,18,19,20,21\},\\
B(1,4) = \{22,23,24,25,26,27,28\},\;\; &B(1,5) = \{29,30,31,32,33,34,35\},\;\; &B(1,6) = \{36,37,38,39,40,41,42\},\\
B(1,7) = \{43,44,45,46,47,48,49\},
&B(2,1) = \{1,8,15,22,29,36,43\}, &B(2,2) = \{2,9,16,23,30,37,44\},\\ B(2,3) = \{3,10,17,24,31,38,45\},
&B(2,4) = \{4,11,18,25,32,39,46\}, &B(2,5) = \{5,12,19,26,33,40,47\},\\ B(2,6) = \{6,13,20,27,34,41,48\},
&B(2,7) = \{7,14,21,28,35,42,49\}.       
\end{IEEEeqnarray*}
%
%\end{example}
%\begin{example}\label{ex-3}
%For the user-to-cache association bipartite graph shown in Fig.~\ref{graph1} 
%
%
Partition sets $\mathcal{P}_1$ and $\mathcal{P}_2$ as following 
\begin{IEEEeqnarray*}{l}
\mathcal{P}_{(i,1)} = \{B(i,1), B(i,2)\}, \mathcal{P}_{(i,2)} = \{B(i,3), B(i,4)\}, \mathcal{P}_{(i,3)} = \{B(i,5), B(i,6), B(i,7)\}  \text{ for } 1\leq i\leq 2.
\end{IEEEeqnarray*}
%We consider the cases when $t \leq b - (z-1)\floor{b/z} = 2$.
%
%\newline\textbf{Case II.1:} \textbf{$t = 1$}. We have $t^\prime = t_z = 1$.
%Let $\frac{M}{N} = \frac{t}{b} = \frac{t}{4}$ where $t \in \mathbb{N}$. 

% The definition of $B_{c(i,j)}$ depends upon the value of $\frac{M}{N}$. %In the following, we following we consider all values of $t$ and show the corresponding placement (decided by $B_{c(i,j)}$) and delivery.
%
%\begin{case}
%$t = 1$.
%\end{case}
\textit{Placement:} Split each file into $b^m = 49$ subfiles. The cache $c(i,j)$ stores the subfiles (of all files) indexed by the elements of the blocks contained in $B_{c(i,j)}$, where for $i = 1,2$,
\begin{IEEEeqnarray*}{ll}
B_{c(i,1)} = B_{c(i,2)} = \{B(i,1),B(i,2)\}, \quad
& B_{c(i,3)} = B_{c(i,4)} = \{B(i,3),B(i,4)\},\\
B_{c(i,5)} = B_{c(i,6)} = \{B(i,5),B(i,6)\},
& B_{c(i,7)} = \{B(i,5), B(i,7)\}.
\end{IEEEeqnarray*}
%
%\end{example} 
%
%\begin{example}\label{ex-4}
%For example~\ref{ex-3} 
%We have the following for $t=1$.
%Let set $B_{k(i,j)}$ denote the set of blocks such that if $B \in B_{k(i,j)}$ then for some cache $c(i^\prime,j^\prime) \in C_{k(i,j)}$ we have $B \in B_{c(i^\prime,j^\prime)}$. 
So we have for $i = 1,2$,
%whose contents are the indices of the subfiles stored in the caches accessed by user $k(i,j)$.
\begin{IEEEeqnarray*}{llll}
B_{k(i,1)} =  \mathcal{P}_i\setminus B(i,7),\quad &B_{k(i,2)} =  \mathcal{P}_i\setminus B(i,7),\quad
&B_{k(i,3)} =  \mathcal{P}_i\setminus B(i,7),\quad &B_{k(i,4)} =  \mathcal{P}_i\setminus B(i,7),\\
B_{k(i,5)} =  \mathcal{P}_i\setminus B(i,7), &B_{k(i,6)} =  \mathcal{P}_i\setminus B(i,7),
&B_{k(i,7)} =  \mathcal{P}_i\setminus B(i,6).
% \text{ for } 1\leq i\leq 2.
\end{IEEEeqnarray*}
%
%\end{example}
%
\begin{figure}[h]
\centering
\includegraphics[width=0.7\textwidth]{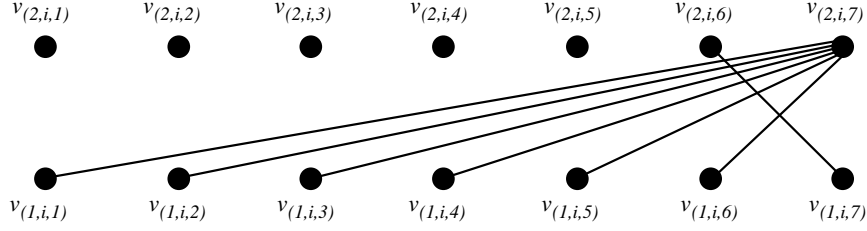}
\caption{Let the graph shown in the figure be denoted by $\bar{J}_i$. The demand graph $\bar{J} = (V_1,V_2,\bar{E})$ for the user-to-cache association bipartite graph $J = (U,C,E)$ considered in the second example in Section~\ref{sec:ex:2} is the disjoint union of $\bar{J}_1$ and $\bar{J}_2$. An edge connects $v_{(1,i,j)}$ to $v_{(2,i^\prime,j^*)}$ if and only if $i = i^\prime$ and the set $B_{f_{M_i}^{-1}(c(i,j))}$ does not contain the block $B(i,j^*)$.}
\label{graphj2}
\end{figure}

%\begin{example}\label{ex-5}

We now construct a new graph that we call as the demand graph $\bar{J} = (V_1,V_2,\bar{E})$, an abstraction of which is shown in Fig.~\ref{graphj2}. We have for $l = 1,2$, $V_l = \{v_{(l,i,j)} | 1\leq i\leq 2, 1\leq j\leq 7\}$. There is a bijection from the vertices in $V_1$ to the caches in $C$, and there is another bijection from the vertices in $V_2$ to the blocks in $\mathcal{P}_1 \cup \mathcal{P}_2$. Say for $i=1,2$, $1\leq j\leq 7$, the vertex $v_{(1,i,j)}$ maps to the cache $c(i,j)$, similarly, say that the vertex $v_{(2,i,j)}$ maps to the block $B(i,j)$. For $i=1,2$, $1\leq j\leq 7$ there exists an edge in $\bar{E}$ connecting vertex $v_{(1,i,j)} \in V_1$ to vertex $v_{(2,\bar{i},j^*)} \in V_2$ if and only if $i = \bar{i}$ and $B(i,j^*) \notin B_{k(i,j^\prime)}$ where $k(i,j^\prime) = f_{M_i}^{-1}(c(i,j))$.

We define a map $f_{(i,j)}: \{1\} \rightarrow \{1,2,3,4,5,6,7\}\setminus\{l \,|\, B(i,l) \in B_{k(i,j)}\} $ for $1\leq i\leq 2$, $1\leq j\leq 7$ as following.
\begin{IEEEeqnarray*}{l}
f_{(i,j)}(1) = 7 \text{ for } i=1,2, 1\leq j\leq 6\\
f_{(i,7)}(1) = 6 \text{ for } i=1,2.
\end{IEEEeqnarray*}
We define a partial matching $M^1_{j_1,j_2}$ for  $1\leq j_1,j_2\leq 7$ as following.
\begin{IEEEeqnarray*}{l}
M^i_{j_1,j_2} = \{(v_{(1,i,j_i)}, v_{(2,i,f_{(i,j_i^\prime)}(1))})\, | \, i=1,2 \} \text{ where }
k(i,j_i^\prime) = f_{M_i}^{-1}(c(i,j_i)).
\end{IEEEeqnarray*}
So we have
%Consider the partial matching $M^1_{j_1,j_2}$ and $M^2_{j_1,j_2}$ for $1\leq j_1,j_2 \leq 4$ shown below (total $32$ matchings). %Note that we use $f_{M_i}$ as shown in Example~\ref{ex-fm}.
\begin{IEEEeqnarray*}{l}
M^1_{j_1,j_2} = \{ (v_{(1,1,j_1)},v_{(2,1,7)}), (v_{(1,2,j_2)},v_{(2,2,7)})\} \text{ for } 1\leq j_1,j_2\leq 6\\
M^1_{j_1,7} = \{ (v_{(1,1,j_1)},v_{(2,1,7)}), (v_{(1,2,7)},v_{(2,2,6)})\} \text{ for } 1\leq j_1\leq 6\\
M^1_{7,j_2} = \{ (v_{(1,1,7)},v_{(2,1,6)}), (v_{(1,2,j_2)},v_{(2,2,7)})\} \text{ for } 1\leq j_1\leq 6\\
M^1_{7,7} = \{ (v_{(1,1,7)},v_{(2,1,6)}), (v_{(1,2,7)},v_{(2,2,6)})\}.
\end{IEEEeqnarray*}
For each matching $M^1_{j_1,j_2}$ for $1\leq j_1,j_2 \leq 7$, we transmit one subfile $Y^1_{j_1,j_2}$. For a matching $M^1_{j_1,j_2} = \newline \{(v_{(1,1,j_1)},v_{(2,1,j_1^*)}), (v_{(1,2,j_2)},v_{(2,2,j_2^*)})\}$, define $S^{1,1}_{j_1,j_2} = B(1,j_1^*) \cap B(2,j_2)$ and $S^{1,2}_{j_1,j_2} = B(1,j_1) \cap B(2,j_2^*)$. We transmit the following subfiles
\begin{IEEEeqnarray*}{l}
Y^1_{j_1,j_2} = W^{d_{f_{M_1}^{-1}(c(1,j_1))}}(S^{1,1}_{j_1,j_2}) + W^{d_{f_{M_2}^{-1}(c(2,j_2))}}(S^{1,2}_{j_1,j_2})\; \text{ for } 1\leq j_1,j_2 \leq 7.
\end{IEEEeqnarray*}
The transmissions $Y^1_{j_1,j_2}$ along with the set $S^{1,i}_{j_1,j_2}$ for $i = 1, 2$, $1\leq j_1,j_2 \leq 7$, are shown in the following.
\begin{IEEEeqnarray*}{lll}
S^{1,1}_{1,1} = B(1,7) \cap B(2,1) = \{43\}, \quad &S^{1,2}_{1,1} = B(1,1) \cap B(2,7) = \{7\},\qquad
&Y^1_{1,1} = W^{d_{(1,1)}}(43) + W^{d_{(2,1)}}(7)\\
S^{1,1}_{1,2} = B(1,7) \cap B(2,2) = \{44\}, &S^{1,2}_{1,2} = B(1,1) \cap B(2,7) = \{7\},
&Y^1_{1,2} = W^{d_{(1,1)}}(44) + W^{d_{(2,2)}}(7)\\
S^{1,1}_{1,3} = B(1,7) \cap B(2,3) = \{45\}, &S^{1,2}_{1,3} = B(1,1) \cap B(2,7) = \{7\},
&Y^1_{1,3} = W^{d_{(1,1)}}(45) + W^{d_{(2,3)}}(7)\\
S^{1,1}_{1,4} = B(1,7) \cap B(2,4) = \{46\}, &S^{1,2}_{1,4} = B(1,1) \cap B(2,7) = \{7\},
&Y^1_{1,4} = W^{d_{(1,1)}}(46) + W^{d_{(2,4)}}(7)\\
S^{1,1}_{1,5} = B(1,7) \cap B(2,5) = \{47\}, &S^{1,2}_{1,5} = B(1,1) \cap B(2,7) = \{7\},
&Y^1_{1,5} = W^{d_{(1,1)}}(47) + W^{d_{(2,5)}}(7)\\
S^{1,1}_{1,6} = B(1,7) \cap B(2,6) = \{48\}, &S^{1,2}_{1,6} = B(1,1) \cap B(2,7) = \{7\},
&Y^1_{1,6} = W^{d_{(1,1)}}(48) + W^{d_{(2,6)}}(7)\\
S^{1,1}_{1,7} = B(1,7) \cap B(2,7) = \{49\}, &S^{1,2}_{1,7} = B(1,1) \cap B(2,6) = \{6\},
&Y^1_{1,7} = W^{d_{(1,1)}}(49) + W^{d_{(2,7)}}(6)\\
S^{1,1}_{2,1} = B(1,7) \cap B(2,1) = \{43\}, &S^{1,2}_{2,1} = B(1,2) \cap B(2,7) = \{14\},
&Y^1_{2,1} = W^{d_{(1,2)}}(43) + W^{d_{(2,1)}}(14)\\
S^{1,1}_{2,2} = B(1,7) \cap B(2,2) = \{44\}, &S^{1,2}_{2,2} = B(1,2) \cap B(2,7) = \{14\},
&Y^1_{2,2} = W^{d_{(1,2)}}(44) + W^{d_{(2,2)}}(14)\\
S^{1,1}_{2,3} = B(1,7) \cap B(2,3) = \{45\}, &S^{1,2}_{2,3} = B(1,2) \cap B(2,7) = \{14\},
&Y^1_{2,3} = W^{d_{(1,2)}}(45) + W^{d_{(2,3)}}(14)\\
S^{1,1}_{2,4} = B(1,7) \cap B(2,4) = \{46\}, &S^{1,2}_{2,4} = B(1,2) \cap B(2,7) = \{14\},
&Y^1_{2,4} = W^{d_{(1,2)}}(46) + W^{d_{(2,4)}}(14)\\
S^{1,1}_{2,5} = B(1,7) \cap B(2,5) = \{47\}, &S^{1,2}_{2,5} = B(1,2) \cap B(2,7) = \{14\},
&Y^1_{2,5} = W^{d_{(1,2)}}(47) + W^{d_{(2,5)}}(14)\\
S^{1,1}_{2,6} = B(1,7) \cap B(2,6) = \{48\}, &S^{1,2}_{2,6} = B(1,2) \cap B(2,7) = \{14\},
&Y^1_{2,6} = W^{d_{(1,2)}}(48) + W^{d_{(2,6)}}(14)\\
S^{1,1}_{2,7} = B(1,7) \cap B(2,7) = \{49\}, &S^{1,2}_{2,7} = B(1,2) \cap B(2,6) = \{13\},
&Y^1_{2,7} = W^{d_{(1,2)}}(48) + W^{d_{(2,7)}}(13)\\
S^{1,1}_{3,1} = B(1,7) \cap B(2,1) = \{43\}, &S^{1,2}_{3,1} = B(1,3) \cap B(2,7) = \{21\},
&Y^1_{3,1} = W^{d_{(1,3)}}(43) + W^{d_{(2,1)}}(21)\\
S^{1,1}_{3,2} = B(1,7) \cap B(2,2) = \{44\}, &S^{1,2}_{3,2} = B(1,3) \cap B(2,7) = \{21\},
&Y^1_{3,2} = W^{d_{(1,3)}}(44) + W^{d_{(2,2)}}(21)\\
S^{1,1}_{3,3} = B(1,7) \cap B(2,3) = \{45\}, &S^{1,2}_{3,3} = B(1,3) \cap B(2,7) = \{21\},
&Y^1_{3,3} = W^{d_{(1,3)}}(45) + W^{d_{(2,3)}}(21)\\
S^{1,1}_{3,4} = B(1,7) \cap B(2,4) = \{46\}, &S^{1,2}_{3,4} = B(1,3) \cap B(2,7) = \{21\},
&Y^1_{3,4} = W^{d_{(1,3)}}(46) + W^{d_{(2,4)}}(21)\\
S^{1,1}_{3,5} = B(1,7) \cap B(2,5) = \{47\}, &S^{1,2}_{3,5} = B(1,3) \cap B(2,7) = \{21\},
&Y^1_{3,5} = W^{d_{(1,3)}}(47) + W^{d_{(2,5)}}(21)\\
S^{1,1}_{3,6} = B(1,7) \cap B(2,6) = \{48\}, &S^{1,2}_{3,6} = B(1,3) \cap B(2,7) = \{21\},
&Y^1_{3,6} = W^{d_{(1,3)}}(48) + W^{d_{(2,6)}}(21)\\
S^{1,1}_{3,7} = B(1,7) \cap B(2,7) = \{49\}, &S^{1,2}_{3,7} = B(1,3) \cap B(2,6) = \{20\},
&Y^1_{3,7} = W^{d_{(1,3)}}(49) + W^{d_{(2,7)}}(20)\\
S^{1,1}_{4,1} = B(1,7) \cap B(2,1) = \{43\}, &S^{1,2}_{4,1} = B(1,4) \cap B(2,7) = \{28\},
&Y^1_{4,1} = W^{d_{(1,4)}}(43) + W^{d_{(2,1)}}(28)\\
S^{1,1}_{4,2} = B(1,7) \cap B(2,2) = \{44\}, &S^{1,2}_{4,2} = B(1,4) \cap B(2,7) = \{28\},
&Y^1_{4,2} = W^{d_{(1,4)}}(44) + W^{d_{(2,2)}}(28)\\
S^{1,1}_{4,3} = B(1,7) \cap B(2,3) = \{45\}, &S^{1,2}_{4,3} = B(1,4) \cap B(2,7) = \{28\},
&Y^1_{4,3} = W^{d_{(1,4)}}(45) + W^{d_{(2,3)}}(28)\\
S^{1,1}_{4,4} = B(1,7) \cap B(2,4) = \{46\}, &S^{1,2}_{4,4} = B(1,4) \cap B(2,7) = \{28\},
&Y^1_{4,4} = W^{d_{(1,4)}}(46) + W^{d_{(2,4)}}(28)\\
S^{1,1}_{4,5} = B(1,7) \cap B(2,5) = \{47\}, &S^{1,2}_{4,5} = B(1,4) \cap B(2,7) = \{28\},
&Y^1_{4,5} = W^{d_{(1,4)}}(47) + W^{d_{(2,5)}}(28)\\
S^{1,1}_{4,6} = B(1,7) \cap B(2,6) = \{48\}, &S^{1,2}_{4,6} = B(1,4) \cap B(2,7) = \{28\},
&Y^1_{4,6} = W^{d_{(1,4)}}(48) + W^{d_{(2,6)}}(28)\\
S^{1,1}_{4,7} = B(1,7) \cap B(2,7) = \{49\}, &S^{1,2}_{4,7} = B(1,4) \cap B(2,6) = \{27\},
&Y^1_{4,7} = W^{d_{(1,4)}}(49) + W^{d_{(2,7)}}(27)\\
S^{1,1}_{5,1} = B(1,7) \cap B(2,1) = \{43\}, &S^{1,2}_{5,1} = B(1,5) \cap B(2,7) = \{35\},
&Y^1_{5,1} = W^{d_{(1,5)}}(43) + W^{d_{(2,1)}}(35)\\
S^{1,1}_{5,2} = B(1,7) \cap B(2,2) = \{44\}, &S^{1,2}_{5,2} = B(1,5) \cap B(2,7) = \{35\},
&Y^1_{5,2} = W^{d_{(1,5)}}(44) + W^{d_{(2,2)}}(35)\\
S^{1,1}_{5,3} = B(1,7) \cap B(2,3) = \{45\}, &S^{1,2}_{5,3} = B(1,5) \cap B(2,7) = \{35\},
&Y^1_{5,3} = W^{d_{(1,5)}}(45) + W^{d_{(2,3)}}(35)\\
S^{1,1}_{5,4} = B(1,7) \cap B(2,4) = \{46\}, &S^{1,2}_{5,4} = B(1,5) \cap B(2,7) = \{35\},
&Y^1_{5,4} = W^{d_{(1,5)}}(46) + W^{d_{(2,4)}}(35)\\
S^{1,1}_{5,5} = B(1,7) \cap B(2,5) = \{47\}, &S^{1,2}_{5,5} = B(1,5) \cap B(2,7) = \{35\},
&Y^1_{5,5} = W^{d_{(1,5)}}(47) + W^{d_{(2,5)}}(35)\\
S^{1,1}_{5,6} = B(1,7) \cap B(2,6) = \{48\}, &S^{1,2}_{5,6} = B(1,5) \cap B(2,7) = \{35\},
&Y^1_{5,6} = W^{d_{(1,5)}}(48) + W^{d_{(2,6)}}(35)\\
S^{1,1}_{5,7} = B(1,7) \cap B(2,7) = \{49\}, &S^{1,2}_{5,7} = B(1,5) \cap B(2,6) = \{34\},
&Y^1_{5,7} = W^{d_{(1,5)}}(49) + W^{d_{(2,7)}}(34)\\
S^{1,1}_{6,1} = B(1,7) \cap B(2,1) = \{43\}, &S^{1,2}_{6,1} = B(1,6) \cap B(2,7) = \{42\},
&Y^1_{6,1} = W^{d_{(1,6)}}(43) + W^{d_{(2,1)}}(42)\\
S^{1,1}_{6,2} = B(1,7) \cap B(2,2) = \{44\}, &S^{1,2}_{6,2} = B(1,6) \cap B(2,7) = \{42\},
&Y^1_{6,2} = W^{d_{(1,6)}}(44) + W^{d_{(2,2)}}(42)\\
S^{1,1}_{6,3} = B(1,7) \cap B(2,3) = \{45\}, &S^{1,2}_{6,3} = B(1,6) \cap B(2,7) = \{42\},
&Y^1_{6,3} = W^{d_{(1,6)}}(45) + W^{d_{(2,3)}}(42)\\
S^{1,1}_{6,4} = B(1,7) \cap B(2,4) = \{46\}, &S^{1,2}_{6,4} = B(1,6) \cap B(2,7) = \{42\},
&Y^1_{6,4} = W^{d_{(1,6)}}(46) + W^{d_{(2,4)}}(42)\\
S^{1,1}_{6,5} = B(1,7) \cap B(2,5) = \{47\}, &S^{1,2}_{6,5} = B(1,6) \cap B(2,7) = \{42\},
&Y^1_{6,5} = W^{d_{(1,6)}}(47) + W^{d_{(2,5)}}(42)\\
S^{1,1}_{6,6} = B(1,7) \cap B(2,6) = \{48\}, &S^{1,2}_{6,6} = B(1,6) \cap B(2,7) = \{42\},
&Y^1_{6,6} = W^{d_{(1,6)}}(48) + W^{d_{(2,6)}}(42)\\
S^{1,1}_{6,7} = B(1,7) \cap B(2,7) = \{49\}, &S^{1,2}_{6,7} = B(1,6) \cap B(2,6) = \{41\},
&Y^1_{6,7} = W^{d_{(1,6)}}(49) + W^{d_{(2,7)}}(41)\\
S^{1,1}_{7,1} = B(1,6) \cap B(2,1) = \{36\}, &S^{1,2}_{7,1} = B(1,7) \cap B(2,7) = \{49\},
&Y^1_{7,1} = W^{d_{(1,7)}}(36) + W^{d_{(2,1)}}(49)\\
S^{1,1}_{7,2} = B(1,6) \cap B(2,2) = \{37\}, &S^{1,2}_{7,2} = B(1,7) \cap B(2,7) = \{49\},
&Y^1_{7,2} = W^{d_{(1,7)}}(37) + W^{d_{(2,2)}}(49)\\
S^{1,1}_{7,3} = B(1,6) \cap B(2,3) = \{38\}, &S^{1,2}_{7,3} = B(1,7) \cap B(2,7) = \{49\},
&Y^1_{7,3} = W^{d_{(1,7)}}(38) + W^{d_{(2,3)}}(49)\\
S^{1,1}_{7,4} = B(1,6) \cap B(2,4) = \{39\}, &S^{1,2}_{7,4} = B(1,7) \cap B(2,7) = \{49\},
&Y^1_{7,4} = W^{d_{(1,7)}}(39) + W^{d_{(2,4)}}(49)\\
S^{1,1}_{7,5} = B(1,6) \cap B(2,5) = \{40\}, &S^{1,2}_{7,5} = B(1,7) \cap B(2,7) = \{49\},
&Y^1_{7,5} = W^{d_{(1,7)}}(40) + W^{d_{(2,5)}}(49)\\
S^{1,1}_{7,6} = B(1,6) \cap B(2,6) = \{41\}, &S^{1,2}_{7,6} = B(1,7) \cap B(2,7) = \{49\},
&Y^1_{7,6} = W^{d_{(1,7)}}(41) + W^{d_{(2,6)}}(49)\\
S^{1,1}_{7,7} = B(1,6) \cap B(2,7) = \{42\}, &S^{1,2}_{7,7} = B(1,7) \cap B(2,6) = \{48\},
&Y^1_{7,7} = W^{d_{(1,7)}}(42) + W^{d_{(2,7)}}(48).
\end{IEEEeqnarray*}
After all the above transmissions, it can be verified that all the users can retrieve its demanded files.

\section{Conclusion}\label{conclusion}
Most works on the MACC problem (which assumes the number of users to be the same as the number of caches) in the literature assumes a cyclic wrap around the user-to-cache association, which has been introduced in \cite{hachem}. A natural question is whether some other user-to-cache association can provide better results. This question has been studied very recently in \cite{brunero} for the MACC problem where the number of users is more than the number of caches (other than the trivial cases). We address this question for the MACC problem, which has the same number of users and caches. We show that the user-to-cache associations considered in this paper can provide better results than all other existing MACC schemes in some aspects.

%other cache associations can perform better than all of the corresponding available schemes

%Hachem \textit{et al.} mentioned in \cite{hachem} that a cyclic wrap around user-to-cache association was assumed for the sake of simplicity. We studied this problem under a more robust user-to-cache association. We provided a MACC scheme that is applicable whenever the user-to-cache association bipartite graph satisfies three graph-theoretic conditions. We also discussed our underlying observation that connects partial matchings of a demand graph and the cliques of the side information graph of the corresponding index coding problem.  One way to extend our work is to find MACC schemes that are applicable to more generalized user-to-cache association bipartite graphs. Specifically, we anticipate that condition \textbf{C2} can be replaced by a condition that allows the sets $C_{(i,1)},C_{(i,2)},\ldots,C_{(i,z)}$ to contain any arbitrary number of vertices. 

%to explore if both the conditions \textbf{C1} and \textbf{C2} can be replaced by a condition on existence of matchings of a bipartite graph. The larger problem to find MACC schemes that work on more generalized user-to-cache association bipartite graphs remains open.
% 
\appendices

\section{Proofs of Lemma~\ref{lemmax1} and Corollary~\ref{coro1}}
\subsection{Proof of Lemma~\ref{lemmax1}}\label{proof:lemmax1}
\begin{IEEEproof}
For $1\leq i\leq m$, $0\leq j\leq b-1$, let $B(i,j)$ be the $j^{\text{th}}$ block in the $i^{\text{th}}$ parallel class for some ordering of the parallel classes and the blocks in each parallel class. For $1\leq n\leq m$, $n\neq i$, $1\leq l_n \leq b$, we have $\cap_{n=1,n\neq i}^{m} B(n,l_n) \cap B(i,j) = \mu_m$.  Furthermore, if $l \neq l^\prime$, then $B(n,l) \cap B(n,l^\prime) = \emptyset$, as blocks in the parallel class $\mathcal{P}_x$ partitions $X$. As a result, for $1\leq n\leq m$, $n\neq i$, $1\leq l_n^\prime \leq b$, if at least for some value of $1\leq x\leq m$, $x\neq i$, we have $l_x \neq l_x^\prime$, then $\{\cap_{n=1,n\neq i}^{m} B(n,l_n) \cap B(i,j)\} \cap \{\cap_{n=1,n \neq i}^{m} B(n,l_n^\prime) \cap B(i,j)\} = \emptyset$. Since for $1\leq n\leq m$, $n\neq i$, $l_n$ can be chose in $b$ ways, there are $b^{m-1}$ instances of the set $\cap_{n=1,n\neq i}^{m} B(n,l_n)$. Furthermore these sets are all disjoint as argued above. With each such set, $B(i,j)$ has $\mu_m$ distinct elements in common. So block $B(i,j)$ must have at least $\mu_m b^{m-1}$ distinct elements. Let $V$ be the set of these $\mu_m b^{m-1}$ elements.

We now show that block $B(i,j)$  does not contain any other element. Let $a \in X$ be an element such that $a \notin V$ but $a \in B(i,j)$. Then in each parallel class $\mathcal{P}_n$, $1\leq n\leq m$, $n\neq i$, there is a block $B(n,l_n)$, where $1\leq l_n \leq b $, such that $a \in B(n,l_n)$ (as blocks in $\mathcal{P}_n$ partitions $X$). So $a \in \cap_{n=1, n\leq i}^{m} B(n,l_n) \cap B(i,j)$. Then $a$ must belong to the set $V$. This contradicts the assumption that $a \notin V$.

Since there are $b$ blocks in each parallel class and each block has $\mu_m b^{m-1}$ elements, $X$ must have exactly $\mu_m b^{m}$ elements.
\end{IEEEproof}

\subsection{Proof of Corollary~\ref{coro1}}\label{proof:coro1}
\begin{IEEEproof}
Even though the proof this corollary is contained in the proof Lemma~\ref{lemmax1}, we provide an alternate proof here. Say for any positive integer $m$, $[m]_i = \{1,2,\ldots,m\} \setminus \{i\}$. Let $[x]_i \subseteq [m]_i$ where $x \leq m$. Then, $I([x]_i,j) = I(\{l_y | y \in [x]_i \} \cup j)$. We show the following results holds: $I([x]_i,j) = I([x-1]_i,j)$.
%%I_{[x]} = I_{x-1}
%%\end{IEEEeqnarray}
\begin{IEEEeqnarray*}{l}
\cup_{\forall y\in [x]} \cup_{l_y=1}^b I([x]_i,j)\\
= \cup_{\forall y\in [x-1]} \cup_{l_y=1}^b \cup_{l_x = 1}^b \{I([x-1]_i,j)\cap B(x,l_x)\}\\ 
= \cup_{\forall y\in [x-1]} \cup_{l_y=1}^b \{I([x-1]_i,j)\cap B(x,1)\}\cup \{I([x-1]_i,j)\cap B(x,2)\} \cup \cdots \cup \{I([x-1]_i,j)\cap B(x,m)\}\\
= \cup_{\forall y\in [x-1]} \cup_{l_y=1}^b I([x-1]_i,j)\cap \{B(x,1)\cup B(x,2) \cup \cdots \cup B(x,m)\}\\
=  \cup_{\forall y\in [x-1]} \cup_{l_y=1}^b I([x-1]_i,j)\cap X\\
= \cup_{\forall y\in [x-1]} \cup_{l_y=1}^b I([x-1]_i,j).
\end{IEEEeqnarray*}
Similarly, it can be shown that $I([x-1]_i,j) = I([x-2]_i,j)$. And hence $I([x]_i,j) = I([x-2]_i,j)$. Thus, proceeding in the same manner would yield $I([x]_i,j) = I([x - (x-1)]_i,j)$. For some $w \in [x]_i$, we have $I([x - (x-1)]_i,j) = I(w,j)$. Now,
\begin{IEEEeqnarray*}{ll}
\cup_{l_w=1}^b I(l_w,j) = B(i,j) \cap \{B(w,1)\cup B(w,2) \cup \cdots \cup B(w,m)\} = B(i,j) \cap X = B(i,j)
\end{IEEEeqnarray*}
\end{IEEEproof}
\section{Proof of Theorem~\ref{theorem-1}}\label{sysMCRD}
%\section{Systematic construction of MCRDs}\label{sysMCRD}
%In this section we provide a systematic method to construct MCRDs.
In reference~\cite{gen_crd} the authors show a new class of CRD. A special case of these CRDs result in MCRD. These MCRDs has the same parameters as that claimed in this theorem. So there already exists a proof for the theorem. In this section we provide an alternate proof. We remark that we have obtained our proof independently to \cite{gen_crd}. We also state that our proof is motivated by the work of Tang and Ramamoorthy in \cite{rama}.
 
For any positive integer $q$, let $\mathbb{Z}_q$ be the ring of integers modulo $q$. Let $\mathbb{Z}_q^m$ be the set of all $m$-length column vectors over $\mathbb{Z}_q$. It can be seen that $\mathbb{Z}_q^m$ has total $q^m$ distinct column vectors. Create a matrix $M_{\{m,q\}}$ by listing all of these $q^m$ distinct column vectors (in any arbitrary order) as columns of $M_{\{m,q\}}$. Hence $M_{\{m,q\}}$ is an $m \times q^m$ matrix. For any matrix $M$, let $M(i,j)$ denote the value at the $i^{\text{th}}$ row and $j^{\text{th}}$ column of $M$ (with the row and column indices starting from $1$ and not $0$). 
For $1\leq i\leq m$, $l \in \mathbb{Z}_q$ we define the following $mq$ sets.
\begin{IEEEeqnarray*}{l}
B(i,l+1) = \{ j \, |\, j \in \{1,2,\ldots, q^m\}\, ,\, M_{m,q}(i,j) = l\}.
\end{IEEEeqnarray*}
In words, $B(i,l+1)$ contains all the column indices $j$ such that the value of $M_{\{m,q\}}$ at the $i^{\text{th}}$ row and $j^{\text{th}}$ column is $l$.
\begin{lemma}\label{01}
For any $1\leq i \leq m$, and $l \in \mathbb{Z}_q$, set $B(i,l+1)$ contains $q^{m-1}$ elements, \textit{i.e.} $|B(i,l+1)| = q^{m-1}$. 
\end{lemma}
\begin{IEEEproof}
First note $\mathbb{Z}_q^m$ is group under vector addition. Since all distinct vectors of $\mathbb{Z}_q^m$ are present as a column of $M_{\{m,q\}}$, the columns of $M_{\{m,q\}}$ form a group. Let $V^{l}_i$ be the set of all columns of $M_{\{m,q\}}$ such that the value at the $i^{\text{th}}$ row of all columns in $V^l_i$ is $l \in \mathbb{Z}_q$. It can be seen that $V^0_i$ is a proper subgroup of $\mathbb{Z}_q^m$. Furthermore, $V^0_i$ has $q$ cosets namely $V^0_i, V^1_i, V^2_i, \ldots , V^{q-1}_i$. And, any of the $q^m$ column vectors of $M_{\{m,q\}}$ must be contained in one of the $q$ cosets $V^l_i$ for $l = 0,1,\ldots, q-1$. As all cosets are of same size, each coset must have $q^{m-1}$ column vectors. 

As $B(i,l+1)$ contains all the column indices of the columns present in $V^l_i$, it is immediate that $|B(i,l+1)| = q^{m-1}$.
\end{IEEEproof}
From $M_{\{m,q\}}$, we construct a design $(X,\mathcal{A})$ where $X = \{1,2,\ldots,q^m\}$, and $\mathcal{A} = \{B(i,l+1) \, | \, 1\leq i\leq m, l \in \mathbb{Z}_q\}$. 
\begin{lemma}\label{02}
$(X,\mathcal{A})$ is a resolvable design.
\end{lemma}
\begin{IEEEproof}
We fist define $m$ parallel classes of $(X,\mathcal{A})$. For $1\leq i\leq m$, let $\mathcal{P}_i = \{B(i,l+1) \,|\, l \in \mathbb{Z}_q\}$. Since the $i^{\text{th}}$ row of every column must contain one value from $\mathbb{Z}_q$, each column index in $\{1,2,\ldots,q^m \}$ must be present in exactly one set contained in $\mathcal{P}_i$. Thus $\mathcal{P}_i$ partitions $X$.

Furthermore, as for $1\leq i\leq m$, $l \in \mathbb{Z}_q$, $B(i,l+1) \in \mathcal{P}_i$, the sets $\{\mathcal{P}_1, \mathcal{P}_2, \ldots, \mathcal{P}_m\}$ partitions $\mathcal{A}$. So $(X,\mathcal{A})$ is a resolvable design.
\end{IEEEproof}

\begin{lemma}\label{03}
$(X,\mathcal{A})$ is a maximal cross resolvable design (MCRD) with $\mu_m = 1$.
\end{lemma}
\begin{IEEEproof}
We have already shown that $(X,\mathcal{A})$ is a resolvable design. Say $l_1,l_2,\ldots,l_m \in \mathbb{Z}_q$ be any $m$ values from $\mathbb{Z}_q$. We show that $|\cap_{i=1}^m B(i,l_i+1)| = 1$.

Since every $m$-length vector over $\mathbb{Z}_q$ appear exactly once as a column vector of $M_{\{m,q\}}$, for some $c \in  \{1,2,\ldots,q^m\}$, the column vector at the $c^{\text{th}}$ column is $\begin{bmatrix} l_1 & l_2 & \cdots & l_m \end{bmatrix}^T$. Hence $c \in B(i,l_i+1)$ for $1\leq i\leq m$, and as every vector occurs exactly once, there $\nexists\, c^\prime \in \{1,2,\ldots,q^m\}$, $c^\prime \neq c$, and $c^\prime \in B(i,l_i+1)$ for $1\leq i\leq m$. So $\cap_{i=1}^m B(i,l_i+1) = c$, and the lemma is proved.
\end{IEEEproof}
\begin{example}\label{ex01}
\begin{IEEEeqnarray*}{l}
M_{\{3,2\}} = \begin{bmatrix}
0 & 0 & 0 & 0 & 1 & 1 & 1 & 1\\
0 & 0 & 1 & 1 & 0 & 0 & 1 & 1\\
0 & 1 & 0 & 1 & 0 & 1 & 0 & 1
\end{bmatrix}
\end{IEEEeqnarray*}
We have $B(1,1) = \{1,2,3,4\}, B(1,2) = \{5,6,7,8\}, B(2,1) = \{1,2,5,6\}, B(2,2) = \{3,4,7,8\}$, \newline $B(3,1) = \{1,3,5,7\}, B(3,2) = \{2,4,6,8 \}$. We now show the parallel classes.
\begin{IEEEeqnarray*}{l}
\mathcal{P}_1 = \{\{1,2,3,4\}, \{5,6,7,8\}\}\\
\mathcal{P}_2 = \{\{1,2,5,6\}, \{3,4,7,8\}\}\\
\mathcal{P}_3 = \{\{1,3,5,7\}, \{2,4,6,8\}\}. 
\end{IEEEeqnarray*}
For $X = \{1,2,\ldots,8\}$, $\mathcal{A} = \{\mathcal{P}_1, \mathcal{P}_2, \mathcal{P}_3\}$, it can be seen that $(X,\mathcal{A})$ is an MCRD with $\mu_3 = 1$.
\end{example}
\begin{example}\label{ex02}
\begin{IEEEeqnarray*}{l}
M_{\{4,2\}} = \begin{bmatrix}
0 & 0 & 0 & 0 & 0 & 0 & 0 & 0 & 1 & 1 & 1 & 1 & 1 & 1 & 1 & 1\\
0 & 0 & 0 & 0 & 1 & 1 & 1 & 1 & 0 & 0 & 0 & 0 & 1 & 1 & 1 & 1\\
0 & 0 & 1 & 1 & 0 & 0 & 1 & 1 & 0 & 0 & 1 & 1 & 0 & 0 & 1 & 1\\
0 & 1 & 0 & 1 & 0 & 1 & 0 & 1 & 0 & 1 & 0 & 1 & 0 & 1 & 0 & 1
\end{bmatrix}
\end{IEEEeqnarray*}
We have $B(1,1) = \{1,2,3,4,5,6,7,8\}$, $B(1,2) = \{9,10,11,12,13,14,15,16\}$,  $B(2,1) = \{1,2,3,4,9,10,11,12\}$, $B(2,2) = \{5,6,7,8,13,14,15,16\}, B(3,1) = \{1,2,5,6,9,10,13,14\}, B(3,2) = \{3,4,7,8,11,12,15,16\}$,\newline $B(4,1) =~\{1,3,5,7,9,11,13,15\}, B(4,2) = \{2,4,6,8,10,12,14,16\}$. We now show the parallel classes.
\begin{IEEEeqnarray*}{l}
\mathcal{P}_1 = \{\{1,2,3,4,5,6,7,8\}, \{9,10,11,12,13,14,15,16\}\}\\ \mathcal{P}_2 = \{\{1,2,3,4,9,10,11,12\}, \{5,6,7,8,13,14,15,16\}\}\\ \mathcal{P}_3 = \{\{1,2,5,6,9,10,13,14\}, \{3,4,7,8,11,12,15,16\}\}\\ \mathcal{P}_4 = \{\{1,3,5,7,9,11,13,15\}, \{2,4,6,8,10,12,14,16\}\}. 
\end{IEEEeqnarray*}
For $X = \{1,2,\ldots,16\}$, $\mathcal{A} = \{\mathcal{P}_1, \mathcal{P}_2, \mathcal{P}_3, \mathcal{P}_4\}$, it can be seen that $(X,\mathcal{A})$ is an MCRD with $\mu_4 = 1$.
\end{example}
\begin{example}\label{ex03}
\begin{IEEEeqnarray*}{l}
M_{3,3} = 
\begin{bmatrix}
0 & 0 & 0 & 0 & 0 & 0 & 0 & 0 & 0 & 1 & 1 & 1 & 1 & 1 & 1 & 1 & 1 & 1 & 2 & 2 & 2 & 2 & 2 & 2 & 2 & 2 & 2\\
0 & 0 & 0 & 1 & 1 & 1 & 2 & 2 & 2 & 0 & 0 & 0 & 1 & 1 & 1 & 2 & 2 & 2 & 0 & 0 & 0 & 1 & 1 & 1 & 2 & 2 & 2\\
0 & 1 & 2 & 0 & 1 & 2 & 0 & 1 & 2 & 0 & 1 & 2 & 0 & 1 & 2 & 0 & 1 & 2 & 0 & 1 & 2 & 0 & 1 & 2 & 0 & 1 & 2\\                                                                                                                                                                                                                                                                                                                                                                                                                                                                                                                                                                                                                                                                                                                                                                                                                                                                                                                                                                                      
\end{bmatrix}
\end{IEEEeqnarray*}
We have $B(1,1) = \{1,2,3,4,5,6,7,8,9\}$, $B(1,2) = \{10,11,12,13,14,15,16,17,18\}$, $B(1,3) = \{19,20,21,22,\allowbreak 23,24,25,26,27\}$,  $B(2,1) = \{1,2,3,10,11,12,19,20,21\}$, $B(2,2) = \{4,5,6,13,14,15,22,23,24\}, B(2,3) = \{7,8,9,16,17,18,25,26,27\}, B(3,1) = \{1,4,7,10,13,16,19,22,25\}, B(3,2) = \{2,5,8,11,14,17,20,23,26\}$, \newline $B(3,3) = \{3,6,9,12,15,18,21,24,27\}$. We now show the parallel classes.
\begin{IEEEeqnarray*}{l}
\mathcal{P}_1 = \{\{1,2,3,4,5,6,7,8,9\}, \{10,11,12,13,14,15,16,17,18\}, \{19,20,21,22,\allowbreak 23,24,25,26,27\}\}\\ \mathcal{P}_2 = \{\{1,2,3,10,11,12,19,20,21\}, \{4,5,6,13,14,15,22,23,24\}, \{7,8,9,16,17,18,25,26,27\}\}\\ \mathcal{P}_3 = \{\{1,4,7,10,13,16,19,22,25\}, \{2,5,8,11,14,17,20,23,26\}, \{3,6,9,12,15,18,21,24,27\}\}. 
\end{IEEEeqnarray*}
For $X = \{1,2,\ldots, 27\}$, $\mathcal{A} = \{\mathcal{P}_1, \mathcal{P}_2, \mathcal{P}_3\}$, it can be seen that $(X,\mathcal{A})$ is an MCRD with $\mu_3 = 1$.
\end{example}
\begin{example}\label{ex04}
\begin{IEEEeqnarray*}{l}
M_{\{2,4\}} = \begin{bmatrix}
0 & 0 & 0 & 0 & 1 & 1 & 1 & 1 & 2 & 2 & 2 & 2 & 3 & 3 & 3 & 3\\
0 & 1 & 2 & 3 & 0 & 1 & 2 & 3 & 0 & 1 & 2 & 3 & 0 & 1 & 2 & 3 
\end{bmatrix}
\end{IEEEeqnarray*}
We have $B(1,1) = \{1,2,3,4\}, B(1,2) = \{5,6,7,8\}, B(1,3) = \{9,10,11,12\}, B(1,4) = \{13,14,15,16\}$, $B(2,1) = \{1,5,9,13\}, B(2,2) = \{2,6,10,14\}, B(2,3) = \{3,7,11,15\}, B(2,4) = \{4,8,12,16\}$. We now show the parallel classes.
\begin{IEEEeqnarray*}{l}
\mathcal{P}_1 = \{\{1,2,3,4\}, \{5,6,7,8\}, \{9,10,11,12\}, \{13,14,15,16\}\}\\ \mathcal{P}_2 = \{\{1,5,9,13\}, \{2,6,10,14\}, \{3,7,11,15\}, \{4,8,12,16\}\}. 
\end{IEEEeqnarray*}
For $X = \{1,2,\ldots,16\}$, $\mathcal{A} = \{\mathcal{P}_1, \mathcal{P}_2\}$, it can be seen that $(X,\mathcal{A})$ is an MCRD with $\mu_2 = 1$. We considered the same design in Example~\ref{ex-a3}.
\end{example}

We now construct an MCRD which has $\mu_m = n$, where $n \geq 1$. Construct a matrix $M_{\{m,q\}}^n$ where each vector belonging to $\mathbb{Z}_q^m$ repeats in exactly $n$ columns (in any arbitrary order). Hence $M_{\{m,q\}}^n$ is an $m \times nq^m$ matrix. For $1\leq i\leq m$, $l \in \mathbb{Z}_q$ we re-define $B(i,l+1)$.
\begin{IEEEeqnarray*}{l}
B(i,l+1) = \{ j \, |\, j \in \{1,2,\ldots, nq^m\}\, ,\, M_{m,q}^n(i,j) = l\}
\end{IEEEeqnarray*}
The following lemma can be proven proceeding similarly to Lemma~\ref{01}.
\begin{lemma}%\label{01}
For any $1\leq i \leq m$, and $l \in \mathbb{Z}_q$, set $B(i,l+1)$ contains $nq^{m-1}$ elements, \textit{i.e.} $|B(i,l+1)| = nq^{m-1}$. 
\end{lemma}
From $M_{\{m,q\}}^n$, we construct a design $(X,\mathcal{A})$ where $X = \{1,2,\ldots,nq^m\}$, and $\mathcal{A} = \{B(i,l+1) \, | \, 1\leq i\leq m, l \in \mathbb{Z}_q\}$. Similar to Lemmas~\ref{02} and \ref{03} it can be shown that $(X,\mathcal{A})$ is an MCRD with $\mu_m = n$.

\begin{example}\label{ex05}
\begin{IEEEeqnarray*}{l}
M_{\{3,2\}}^2 = \begin{bmatrix}
0 & 0 & 0 & 0 & 0 & 0 & 0 & 0 & 1 & 1 & 1 & 1 & 1 & 1 & 1 & 1\\
0 & 0 & 0 & 0 & 1 & 1 & 1 & 1 & 0 & 0 & 0 & 0 & 1 & 1 & 1 & 1\\
0 & 0 & 1 & 1 & 0 & 0 & 1 & 1 & 0 & 0 & 1 & 1 & 0 & 0 & 1 & 1
\end{bmatrix}
\end{IEEEeqnarray*}
$M_{\{3,2\}}^2$ is the same as the first three rows of $M_{4,2}$ considered in Example~\ref{ex02}. For the same blocks and parallel classes ($\mathcal{P}_4$ does not exists in this case), $(X,\mathcal{A})$ is an MCRD with $\mu_3 = 2$, where $X = \{1,2,\ldots,16\}$, $\mathcal{A} = \{\mathcal{P}_1, \mathcal{P}_2, \mathcal{P}_3\}$.
\end{example}

\section{Proofs of Lemmas \ref{P:RK}, \ref{P:SICPS}, \ref{lemmasr1}, \ref{P:SR2}, \ref{P:MR}}\label{compare-proof}
\subsection{Proof of Lemma~\ref{P:RK}}\label{p:RK}
\begin{IEEEproof}
Since $\frac{t}{b} < \frac{1}{b}\floor{\frac{b}{z}}$ our scheme achieves a rate $b - tz$. Then, the values $\frac{M}{N}$ where our scheme achieves a lesser rate that the RK scheme must satisfy
\begin{IEEEeqnarray*}{ll}
&b - tz \leq \frac{(K - mtz)^2}{K}
\implies b - tz \leq \frac{(mb - mtz)^2}{mb}\label{-eq1}\IEEEyesnumber\\
\implies &b - tz \leq \frac{m(b - tz)^2}{b}
\implies 1 \leq \frac{m(b - tz)}{b}\label{-eq2}\IEEEyesnumber\\
\implies &b \leq mb - mtz = K - mtz
\implies t \leq \frac{K - b}{mz}
\implies \frac{t}{b} \leq \frac{K - b}{Kz}.
%\implies &\frac{b}{K} \leq 1 - z\frac{M}{N}\\
%\implies &\frac{M}{N} 
%\leq \frac{1 - \frac{b}{K}}{z} = \frac{1 - \frac{1}{m}}{z}
\end{IEEEeqnarray*}
In equation~(\ref{-eq1}) $t^\prime$ has been replaced by $mt$ as $t/b = t^\prime/K$ (see Table~\ref{table3}). Since $\frac{M}{N} = \frac{t}{b}$, we have $\frac{t}{b} < \frac{1}{b}\floor{\frac{b}{z}}$, which implies $\frac{t}{b} < \frac{1}{z}$, or $b - tz \neq 0$, and hence equation~(\ref{-eq2}) holds.
\end{IEEEproof}

\subsection{Proof of Lemma~\ref{P:SICPS}}\label{p:SICPS}
\begin{IEEEproof}
The subpacketization level of the RK scheme and the SICPS scheme are equal. At $\frac{M}{N} = \frac{1}{b} = \frac{t^\prime}{K}$ the SICPS scheme achieves a subapcketization level $\frac{K}{t^\prime}\binom{K - t^\prime z + t^\prime - 1}{t^\prime - 1} = b\binom{mb - mz + m - 1}{m - 1}$. % We know %Let $n = mb - mz + m - 1$, $r = m - 1$.

Let $n$ and $r$ be positive integers where $n \geq r$.
\begin{IEEEeqnarray*}{ll}
\binom{n}{r} = \frac{n(n-1)(n-2)\cdots(n-(r-1))}{r(r-1)(r-2)\cdots (r - (r-1))} = \frac{n}{r}\frac{(n-1)}{(r-1)}\frac{(n-2)}{(r-2)}\cdots\frac{(n-(r-1))}{(r - (r-1))}\\
\geq \frac{n}{r}\frac{n}{r}\frac{n}{r}\cdots\frac{n}{r} = (\frac{n}{r})^r.\IEEEyesnumber\label{uk1}
\end{IEEEeqnarray*}
Since $b \geq z$, we have $mb - mz + m - 1 \geq m-1$. Then, using equation~(\ref{uk1}) we have
\begin{IEEEeqnarray*}{ll}
b\binom{mb - mz + m - 1}{m - 1} = b(\frac{mb - mz + m - 1}{m - 1})^{m-1}.
\end{IEEEeqnarray*}
Our scheme is guaranteed to achieve a lesser subpacketization than the SICPS scheme when
\begin{IEEEeqnarray*}{ll}
& b^m \leq b(\frac{mb - mz + m - 1}{m - 1})^{m-1}
\implies  b^{m-1} \leq (\frac{mb - mz + m - 1}{m - 1})^{m-1}\\
\implies & b \leq \frac{mb - mz + m - 1}{m - 1}
\implies  b(m-1) \leq mb - mz + m - 1
\implies  mb - b \leq mb - mz + m - 1\\
\implies & b \geq mz - m + 1 = m(z - 1) + 1
\implies  b \geq \frac{K}{b}(z - 1) + 1.\IEEEyesnumber\label{uk2}
\end{IEEEeqnarray*}
Now, 
\begin{IEEEeqnarray*}{ll}
&\sqrt{K(z-1)} + 1 \geq \sqrt{K(z-1)} \implies 1 \geq \frac{\sqrt{K(z-1)}}{\sqrt{K(z-1)} + 1}\\
\implies &\sqrt{K(z-1)} \geq \frac{K(z-1)}{\sqrt{K(z-1)} + 1}
\implies \sqrt{K(z-1)} + 1 \geq \frac{K(z-1)}{\sqrt{K(z-1)} + 1} + 1.
\end{IEEEeqnarray*}
So $b = \sqrt{K(z-1)} + 1$ satisfies equation~(\ref{uk2}). As the right hand side of equation~(\ref{uk2}) decreases with increasing $b$, all values of $b$ greater than or equal to $\sqrt{K(z-1)} + 1$ (where $b$ must divide $K$ as well) must satisfy equation~(\ref{uk2}).

Now, we know that $\binom{n}{r} \geq \binom{n-1}{r-1}$. So the subpacketization level of the NT scheme is always at least as much as the SCIPS scheme. This completes the proof of the lemma.
\end{IEEEproof}

\subsection{Proof of Lemma~\ref{lemmasr1}}\label{p:SR1}
\begin{IEEEproof}
It can be seen that for our scheme the corner points $(\frac{1}{b_1}, b_1 - z)$ and $(\frac{1}{b_2}, b_2 - z)$ are achievable. So any point $(x,y)$ on the straight line connecting the two points is also achievable. We have
\begin{IEEEeqnarray}{l}
y = -b_1b_2x + b_2 + b_1 - z.\label{eqn--5}
\end{IEEEeqnarray}
Let $(x,y)$ be the point such that 
\begin{IEEEeqnarray}{l}
x = \frac{1}{b_1} + \lambda(\frac{1}{b_2} - \frac{1}{b_1}) = \frac{m_1 + \lambda(m_2 - m_1)}{K} = \frac{t^{\prime\prime}}{K}.\label{eqn--2}
\end{IEEEeqnarray}
Since $m_1$ and $m_2$ are such that $gcd(t^{\prime\prime},K) = 1$, at $\frac{M}{N} = x$ the SR1 scheme has a non-trivial achievable corner point.

%We reproduce the rate achieved by the SR1 scheme from \cite{shanuja}. 
%The case $K - t^{\prime\prime} z = 1$ never arises due to the conditions set in the statement of the lemma. %If $K - t^{\prime\prime}z$ is even, then the SR1 scheme achieves a rate

It can be seen from \cite{shanuja} that if $K - t^{\prime\prime}z$ is even, then the SR1 scheme achieves a rate
\begin{IEEEeqnarray*}{l}
R_{SR1} = \sum_{r=\frac{K - t^{\prime\prime}z + 2}{2}}^{K - t^{\prime\prime}z} \frac{2}{1 + \ceil{\frac{t^{\prime\prime}z}{r}}}.
\end{IEEEeqnarray*}
And if $K - t^{\prime\prime}z > 1$ is odd (the case $K - t^{\prime\prime} z = 1$ never arise due to the conditions set in the statement of the lemma), then the SR1 scheme achieves a rate
\begin{IEEEeqnarray*}{l}
R_{SR1} = \frac{1}{1 + \ceil{\frac{2t^{\prime\prime}z}{K - t^{\prime\prime}z + 1}}} +  \sum_{r=\frac{K - t^{\prime\prime}z + 3}{2}}^{K - t^{\prime\prime}z} \frac{2}{1 + \ceil{\frac{t^{\prime\prime}z}{r}}}.
\end{IEEEeqnarray*}
So irrespective of whether $K - t^{\prime\prime}z > 1$ is odd or even, we have 
\begin{IEEEeqnarray}{l}
R_{SR1} \geq \sum_{r=\frac{K - t^{\prime\prime}z + 2}{2}}^{K - t^{\prime\prime}z} \frac{2}{1 + \ceil{\frac{t^{\prime\prime}z}{r}}}
\geq\sum_{r=\frac{K - t^{\prime\prime}z + 2}{2}}^{K - t^{\prime\prime}z} \frac{2}{1 + \frac{t^{\prime\prime}z}{r} + 1} = (K - t^{\prime\prime}z)\frac{2}{2 + \frac{t^{\prime\prime}z}{\frac{K - t^{\prime\prime}z + 2}{2}}} = \frac{(K - t^{\prime\prime}z)(K - t^{\prime\prime}z + 2)}{K+2}.\IEEEeqnarraynumspace\label{eqn--1}
\end{IEEEeqnarray}
Equation~(\ref{eqn--1}) has also been shown to hold in reference \cite{cheng}. So for our scheme achieves a lesser rate at least when
\begin{IEEEeqnarray*}{ll}
& y \leq \frac{(K - t^{\prime\prime}z)(K - t^{\prime\prime}z + 2)}{K+2}\\
\implies & -b_1b_2x + b_2 + b_1 - z \leq \frac{(K - t^{\prime\prime}z)(K - t^{\prime\prime}z + 2)}{K+2}\IEEEyesnumber\label{eq--4}\\
\implies & -b_2 - \lambda(b_1 - b_2) + b_2 + b_1 - z = b_1 - \lambda(b_1 - b_2) - z \leq \frac{(K - t^{\prime\prime}z)(K - t^{\prime\prime}z + 2)}{K+2}.\IEEEyesnumber\label{eq--6}
\end{IEEEeqnarray*}
Equations~(\ref{eq--4}) and (\ref{eq--6}) hold due to equations~(\ref{eqn--5}) and (\ref{eqn--2}) respectively. 
\end{IEEEproof}

\subsection{Proof of Lemma~\ref{P:SR2}}\label{p:SR2}
\begin{IEEEproof}
The rate achieved by the SR2 scheme is 
\begin{IEEEeqnarray*}{l}
\frac{(mb - mtz)(mb - mtz + mt)}{2mb} = \frac{(b-tz)(mb - mtz + mt)}{2b}.
\end{IEEEeqnarray*}
For $\frac{M}{N} \leq \frac{1}{b}\floor(\frac{b}{z})$ our scheme achieves a lesser rate if
\begin{IEEEeqnarray*}{ll}
&b - tz \leq \frac{(b-tz)(mb - mtz + mt)}{2b}\\
\implies &2b \leq mb - mt(z-1) \implies t \leq \frac{mb - 2b}{m(z-1)}\\
\implies &\frac{t}{b} \leq \frac{m - 2}{m(z-1)}.
\end{IEEEeqnarray*}
\end{IEEEproof}

\subsection{Proof of Lemma~\ref{P:MR}}\label{p:MR}
For $z < \frac{K}{2}$, the MR scheme achieves a rate $\frac{1}{K}\ceil{\frac{K(K -z)}{2}}$ files at $\frac{M}{N} = \frac{1}{K}$. Trivially the point $(\frac{1}{K}\ceil{\frac{K}{z}},0)$ is also achievable. It can be seen that the line connecting these two achievable points is above the line connecting the points $(\frac{1}{K},\frac{K-z}{2})$ and $(\frac{1}{z},0)$. If the point $(x,y)$ is on the line connecting the latter two points then $2y = - Kz x + K$. So at $\frac{M}{N} = \frac{1}{b}$ for $b < K$ our scheme achieves a lesser rate if
\begin{IEEEeqnarray}{l}
2(b - z) \leq - Kz \frac{1}{b} + K \implies 2b - 2z \leq -mb z\frac{1}{b} + mb \implies z(m - 2) \leq b(m - 2) \implies z \leq. b\label{new1}
\end{IEEEeqnarray}
Equation~(\ref{new1}) holds when $m \neq 2$. Since $z \leq b$ always holds for our scheme, it means that at $\frac{M}{N} = \frac{1}{b}$, for $b < K$ and $m \geq 3$, our scheme achieves a lesser rate.
%The proof of Theorem~\ref{theorem-1} is immediate from this section.

%The following corollary is immediate.
%\begin{corollary}
%For any positive integer $m,n,$ and $q$, there exists an MCRD $(X,\mathcal{A})$ with $X = \{1,2,\ldots,nq^m\}$, $|\mathcal{A}| = mq$, $m$ parallel classes, $q$ blocks in each parallel class, and $\mu_m = n$.
%\end{corollary}
%
%\section{Table of Notations}\label{A:table}
%
\section{Proof of Theorem~\ref{theorem1}}\label{proof-here}

\begin{table}\label{table1}
\caption{Notations and their significance}
\centering
\begin{tabularx}{\textwidth}{|>{\centering$}p{1cm}<{$} | >{\centering$}p{2.67cm}<{$} | >{\raggedright\arraybackslash}X|}
\hline
\text{Notation} & \text{Range} & \hspace{120pt} Significance\\
\hline
m & \mathbb{N} & $mb$ caches, users, partitioned into $m$ disjoint sets of $b$ caches, users respectively; MCRD has $m$ parallel classes, subpacketization level is $b^m$.\\\hline
b & \mathbb{N}  & $mb$ caches, users, partitioned into $m$ disjoint sets of $b$ caches, users respectively; each parallel class of MCRD has $b$ blocks, subpacketization level is $b^m$.\\\hline
z & 1\leq z \leq b & Each user accesses $z$ caches.\\\hline
W^i & 1 \leq i \leq N &  The $N$ files at the server. Each file $W^i$ is split into $b^m$ subfiles: $W^i(1),W^i(2),\ldots, W^i(b^m)$.\\\hline
K_i & 1\leq i\leq m & $mb$ users partitioned into $m$ disjoint sets $K_1,K_2,\ldots,K_m$ each having $b$ users.\\\hline
k(i,j) & {1\leq i\leq m}, {1\leq j\leq b} & $j^{\text{th}}$ user in $K_i$ for some ordering of the users in $K_i$.\\\hline
C_i & {1\leq i\leq m} & $mb$ caches partitioned into $m$ disjoint sets $C_1,C_2,\ldots,C_m$ each having $b$ caches.\\\hline
c(i,j) & {1\leq i\leq m}, {1\leq j\leq b} & $j^{\text{th}}$ cache in $C_i$ for some ordering of the caches in $C_i$.\\\hline
C_{(i,l)} & {1\leq i\leq m}, {1\leq l\leq z} & Set $C_i$ is partitioned into $z$ disjoints sets $C_{(i,1)},C_{(i,2)},\ldots, C_{(i,z)}$. Set $C_{(i,l)}$ for $1\leq l\leq z-1$ contains $\floor{b/z}$ caches. Set $C_{(i,z)}$ contains $b - (z-1)\floor{b/z}$ caches.\\\hline
C_{k(i,j)} & {1\leq i\leq m}, {1\leq j\leq z} & $C_{k(i,j)}$ is the set of caches user $k(i,j)$ accesses. $C_{k(i,j)} \subseteq C_i$. For $1\leq l\leq z$, $|C_{(i,l)} \cap C_{k(i,j)}| = 1$.\\\hline
f_{M_i} & {1\leq i\leq m} & $f_{M_i}: \{1,2,\ldots, b\} \to \{1,2,\ldots,b\}$. The set $\{f_{M_i}(k(i,1)), f_{M_i}(k(i,2)), \ldots, f_{M_i}(k(i,b))\}$ is a system of distinct representatives of the sets $C_{k(i,1)}, C_{k(i,2)}, \ldots, C_{k(i,b)}$ where $f_{M_i}(k(i,j)) \in C_{k(i,j)}$ for $1\leq j\leq b$ and $f_{M_i}(k(i,j_1)) \neq f_{M_i}(k(i,j_2))$.\\\hline
\mathcal{P}_i & {1\leq i\leq m} & We consider MCRDs with $m$ parallel classes denoted by $\mathcal{P}_1,\mathcal{P}_2,\ldots,\mathcal{P}_m$.\\\hline
B(i,j) & {1\leq i\leq m}, {1\leq j\leq b} & Each parallel class $\mathcal{P}_i$ has $b$ blocks $B(i,1), B(i,2), \ldots, B(i,b)$.\\\hline
\mathcal{P}_{(i,l)} & {1\leq i\leq m}, {1\leq l\leq z} & Set $\mathcal{P}_i$ is partitioned into $z$ disjoints sets $\mathcal{P}_{(i,1)},\mathcal{P}_{(i,2)},\ldots, \mathcal{P}_{(i,z)}$. Set $\mathcal{P}_{(i,l)}$ for $1\leq l\leq z-1$ contains $\floor{b/z}$ blocks. Set $\mathcal{P}_{(i,z)}$ contains $b - (z-1)\floor{b/z}$ blocks.\\\hline
t,t^\prime ,t_z & \mathbb{N} & If cache $c \in C_{(i,l)}$ where $1\leq l\leq z-1$, then $c$ stores $t^\prime$ blocks belonging to $P_{(i,l)}$. If cache $c \in C_{(i,z)}$, then $c$ stores $t_z$ blocks belonging to $P_{(i,z)}$. Value of $t$ is provided with the coded caching problem. If $1 \leq t \leq \floor{b/z}$, $t^\prime = t_z = t$. If $\floor{b/z} < t < b - (z-1)\floor{\frac{b}{z}}$, $t^\prime = \floor{b/z}, t_z = t$. If $t \geq b - (z-1)\floor{\frac{b}{z}}$, $t^\prime = \floor{b/z}, t_z = b - (z-1)\floor{\frac{b}{z}}$.\\\hline
B_{c(i,j)} & {1\leq i\leq m}, {1\leq j\leq b} & $B_{c(i,j)} \subseteq \mathcal{P}_i$. During placement, cache $c(i,j)$ stores the subfiles indexed by the blocks contained in $B_{c(i,j)}$. If $c(i,j) \in C_{(i,l)}$ where $1\leq l\leq z-1$, $B_{c(i,j)}$ contains $B(i,j)$ and another $t^\prime -1$ blocks from $\mathcal{P}_{(i,l)}$. If $c(i,j) \in C_{(i,z)}$, $B_{c(i,j)}$ contains $B(i,j)$ another $t_z - 1$ blocks from $\mathcal{P}_{(i,z)}$.\\\hline
B_{k(i,j)} & {1\leq i\leq m}, {1\leq j\leq b} & Set $B_{k(i,j)}$ denotes the set of blocks whose contents are the indices of the subfiles stored in the caches accessed by user $k(i,j)$. Since $k(i,j)$ accesses the caches in $C_{k(i,j)}$, and each cache $c(x,y)$ stores the subfiles indexed by the elements of the blocks contained in $B_{c(x,y)}$, we have $B_{k(i,j)} = \cup_{c(x,y) \in C_{k(i,j)}} B_{c(x,y)}$.\\\hline
%
%\!\!\! {P_{i}\!\!\setminus\! B_{k(i,j)}} & {1\leq i\leq m}, {1\leq j\leq q} & It is the set of blocks that are in $P_i$, but none of the caches that user $k(i,j)$ accesses stores the subfiles of the indices contained in these blocks.\\\hline
%
f_{(i,j)} & {1\leq i\leq m}, {1\leq j\leq b} & $f_{(i,j)}: \{1,2,\ldots, b- t^\prime (z-1) - t_z \} \to \{1,2,\ldots,b \} \setminus \{l \,|\, B(i,l) \in B_{k(i,j)}\}$. $f_{i,j}(\{1,2,\ldots, b- t^\prime (z-1) - t_z \})$ is the set of all integers $l$ such that $B(i,l) \notin B_{k(i,j)}$.\\\hline
%
%I & & $I: \mathbb{Z}^m \to X^{\mu_m}$. It is defined as following. For $1 \leq l_1,l_2,\ldots, l_m \leq b$, $I(l_1,l_2,\ldots,l_m) = B(1,l_1) \cap B(2,l_2) \cap \cdots \cap B(m,l_m)$. \hline
\end{tabularx}
\end{table}

%
%\begin{example}\label{ex-1}
%
We use a lot of notations in the proof. To enhance the readability of the paper, we have listed the most relevant notations in Table~\ref{table1}. 

Without loss of generality (w.l.o.g.) we assume that for $1\leq l\leq z-1$, $C_{(i,l)} = \{c(i,y+1),c(i,y+2),\ldots,c(i,y+\floor{\frac{b}{z}})\}$ where $y = (l-1)\floor{\frac{b}{z}}$, and $C_{(i,z)} = \{c(i,y+1),c(i,y+2),\ldots,c(i,b)\}$ where $y = (z-1)\floor{\frac{b}{z}}$. The reasoning behind this w.l.o.g. statement is that we can always re-label the caches to achieve the above segregation. %The following lemma is immediate. % also implicitly assumes the segregation of $C_i$ is symmetric over $i$. The latter statement is explained with an example in Example~\ref{ex2}.
\begin{lemma}\label{lemma1}
For $1\leq i\leq m, 1\leq j\leq b$, cache $c(i,j) \in C_{(i,n)}$ for $n = min\{\ceil{\frac{j}{x}},z\}$ where $x = \floor{\frac{b}{z}}$. 
\end{lemma}
\begin{IEEEproof}
\begin{case}
$j > (z-1)\floor{\frac{b}{z}}$.
\end{case}
In this case we know that $c(i,j) \in C_{(i,z)}$. As $\frac{j}{x} > z-1$, we have $n = z$, and hence the lemma hold for this case.
\begin{case}
$1 \leq j \leq (z-1)\floor{\frac{b}{z}}$.
\end{case}
%\textbf{Case II:} 
%
If $c(i,j) \in C_{(i,l)}$ for $1\leq l\leq z-1$ then $(l-1)\floor{\frac{b}{z}} < j \leq l\floor{\frac{b}{z}}$. This implies $(l-1) < \frac{j}{x} \leq l$ (as $1\leq z\leq b$). Hence, $\ceil{j/x} = l$.
\end{IEEEproof}

In $G_i = (K_i,C_i,E_i)$, a vertex $u \in K_i$ can choose a vertex in $C_{(i,l)}$ in $\floor{\frac{b}{z}}$ ways if $1\leq l \leq z-1$ and in $b - (z-1)\floor{\frac{b}{z}}$ ways if $l = z$. So total number of possible user-to-cache associations that satisfy conditions \textbf{C1} and \textbf{C2} is $(\floor{\frac{b}{z}}^{z-1}(b - (z-1)\floor{\frac{b}{z}}))^{bm}$.

%\begin{description}
%\item \textbf{[C3]} There is a perfect matching in $G_i$.
%\end{description}

For each user $k(i,j)$ where $1\leq i\leq m$, $1\leq j\leq b$, we use the notation $C_{k(i,j)}$ to denote the set of caches user $k(i,j)$ accesses. Equivalently, $C_{k(i,j)}$ is the set of all vertices in $C_i$ such that there exists an edge in $E_i$ that connects the user $k(i,j)$ to the vertex. Note, condition \textbf{C3} ensures that $f_{M_i}(k(i,j)) \in C_{k(i,j)}$ for $1\leq i\leq m, 1\leq j\leq b$.  

In Section~\ref{example} we show the placement and delivery of the MACC problem whose user-to-cache association is given by the bipartite graph shown in Fig.~\ref{graph1}.

\subsection{Placement}\label{placement}
Let $(X,\mathcal{A})$ be a maximal cross resolvable design (MCRD) with $m$ parallel classes, $b$ blocks in each parallel class, $X = \{1,2,\ldots,b^m\}$, $|\mathcal{A}| = mb$, and $\mu_m = 1$. The existence of such an MCRD is guaranteed by Theorem~\ref{theorem-1}. As per Lemma~\ref{lemmax1}, each block in $\mathcal{A}$ has exactly $b^{m-1}$ elements. Let $\mathcal{P}_1,\mathcal{P}_2,\ldots,\mathcal{P}_m$ be the $m$ parallel classes of $(X,\mathcal{A})$. Let $B(i,1), B(i,2), \ldots, B(i,b)$ be the $b$ blocks contained in the parallel class $\mathcal{P}_i$.

%The central server contains $N$ files denoted by $W^1,$ $W^2,\ldots,$ $W^N$. Each cache stores $M/N$ fraction of each file, where $M/N$ is called as the normalized memory size. We split each file $W_i$ into $b^m$ subfiles $W^i(1), W^i(2), \ldots, W^i(b^m)$. The number $b^m$ is called subpacketization level. The number $j$ within the parentheses of the notation of a subfile $W^i(j)$ is called as the index of the subfile. 
%Each cache stores $M/N$ fraction of each file. 
For $1\leq i\leq m$, $1\leq j\leq b$, for each cache $c(i,j)$, we define a set $B_{c(i,j)} \subseteq \mathcal{P}_i$; cache $c(i,j)$ stores the subfiles $W^n(l)$ for $\forall n \in \{1,2,\ldots,N\}$ and $l \in B(i,s)$ for $\forall B(i,s) \in B_{c(i,j)}$ (\textit{i.e.}, the cache $c(i,j)$ stores all files indexed by the elements of the blocks contained in $B_{c(i,j)}$). Towards describing $B_{c(i,j)}$, we segregate each parallel class in the following way.

We partition the set of blocks in $\mathcal{P}_i$ for $1\leq i\leq m$ into $z$ disjoint subsets $\mathcal{P}_{(i,1)},\mathcal{P}_{(i,2)},\ldots,\mathcal{P}_{(i,z)}$. For $1\leq l\leq z-1$, the subset $\mathcal{P}_{(i,l)}$ contains $\floor{\frac{b}{z}}$ blocks. Whereas, the subset $\mathcal{P}_{(i,z)}$ contains $b - (z-1)\floor{\frac{b}{z}}$ blocks, \textit{i.e.}, the blocks contained in $\mathcal{P}_{(i,z)}$ are the blocks that are in $\mathcal{P}_i$, but not in the sets $\mathcal{P}_{(i,l)}$ for $1\leq l\leq z-1$. Without loss of generality (w.l.o.g.) we assume that, for $1\leq l\leq z-1$, $\mathcal{P}_{(i,l)} = \{B(i,y+1),B(i,y+2),\ldots,B(i,y+\floor{\frac{b}{z}}\}$ where $y = (l-1)\floor{\frac{b}{z}}$, and $\mathcal{P}_{(i,z)} = \{B(i,y+1),B(i,y+2),\ldots,B(i,b)\}$ where $y = (z-1)\floor{\frac{b}{z}}$. 
Similar to Lemma~\ref{lemma1} we have the following.
\begin{lemma}\label{lemma3}
For $1\leq i\leq m, 1\leq j\leq b$, cache $B(i,j) \in \mathcal{P}_{(i,n)}$ if and only if $n = min\{\ceil{\frac{j}{x}},z\}$ where $x = \floor{\frac{b}{z}}$. 
\end{lemma}

We define two variables $t^\prime,t_z$. If $1 \leq t \leq \floor{b/z}$, $t^\prime = t_z = t$. If $\floor{b/z} < t < b - (z-1)\floor{\frac{b}{z}}$, $t^\prime = \floor{b/z}, t_z = t$. If $t \geq b - (z-1)\floor{\frac{b}{z}}$, $t^\prime = \floor{b/z}, t_z = b - (z-1)\floor{\frac{b}{z}}$. % It will be soon evident from equation~\ref{case1} that $t^\prime = |\mathcal{P}_{(i,l)} \cap B_{c(i,j)}|$ for any $1\leq l\leq z-1$ and $t_z = |\mathcal{P}_{(i,z)} \cap B_{c(i,j)}|$.

%Since $z\ceil{b/z} \geq b$, $tz < b$ implies $t < \ceil{b/z}$. However, when $tz \geq b$, we have $t \geq \ceil{b/z}$. In the former case the following placement runs as it is, while in the latter case, the following placement is to be run fixing $t = \ceil{b/z}$.  

Due to Lemma~\ref{lemma1} and Lemma~\ref{lemma3}, for $1\leq i\leq m$, $1\leq j\leq b$, if $c(i,j) \in C_{(i,l)}$, then $B(i,j) \in \mathcal{P}_{(i,l)}$. 

We now define $B_{c(i,j)}$. For $1\leq i\leq m$, $1\leq j\leq b$, if $c(i,j) \in C_{(i,l)}$ where $1\leq l\leq z-1$, $B_{c(i,j)}$ contains $B(i,j)$ in addition to another $t^\prime -1$ arbitrarily chosen blocks from $\mathcal{P}_{(i,l)}$; if $c(i,j) \in C_{(i,z)}$,  $B_{c(i,j)}$ contains $B(i,j)$ in addition to another $t_z - 1$ arbitrarily chosen blocks from $\mathcal{P}_{(i,z)}$. % Due to Lemma~\ref{lemma1} we have the following.
%
%
%
%\begin{equation}\label{cases1}
\begin{IEEEeqnarray}{l}
B_{c(i,j)} = 
\begin{cases}
B(i,j) \cup \{B(i,j_1), B(i,j_2),\ldots, B(i,j_{t^\prime -1})\}\subseteq \mathcal{P}_{(i,n)} \text{ where } n = \ceil{j/\floor{\frac{b}{z}}}, j\notin \{j_1,j_2\ldots ,j_{t^\prime -1} \} \\\hfill \text{if } \ceil{j/\floor{\frac{b}{z}}} < z.\IEEEeqnarraynumspace\\
B(i,j) \cup  \{B(i,j_1), B(i,j_2),\ldots, B(i,j_{t_z -1})\} \subseteq \mathcal{P}_{(i,z)}, j\notin \{j_1,j_2\ldots,j_{t_z-1}\}  \text{ if } \ceil{j/\floor{\frac{b}{z}}} \geq z.
\end{cases}\label{case1}
\end{IEEEeqnarray}
%\end{equation}
Because each block contains $b^{m-1}$ subfiles, if $c(i,j) \in C_{(i,l)}$ where $1\leq l\leq z-1$ then $c(i,j)$ stores $b^{m-1}Nt^\prime$ subfiles, and if $c(i,j) \in C_{(i,z)}$ then $c(i,j)$ stores $b^{m-1}Nt_z$ subfiles. Since $t^\prime, t_z \leq t$ and that each cache can store upto $M = \frac{tN}{b}$ files, which is equivalent to $M b^m$ subfiles or $b^{m-1} N t$ subfiles, the placed content does not exceed available cache memory size. 

%As  and $M = \frac{tN}{•}$
%
%
We have the following lemma as a result of $B_{c(i,j)}$.
\begin{lemma}\label{lemma2}
For any $1\leq i\leq m$, $1\leq j_1,j_2 \leq b$, $1\leq l_1,l_2\leq z$, $l_1 \neq l_2$, if $c(i,j_1) \in C_{(i,l_1)}$ and $c(i,j_2) \in C_{(i,l_2)}$, then $B_{c(i,j_1)} \cap B_{c(i,j_2)} = \emptyset$. 
\end{lemma}
\begin{IEEEproof}
Since $l_1 \neq l_2$ we have $\mathcal{P}_{(i,l_1)} \neq \mathcal{P}_{(i,l_2)}$, and from Lemma~\ref{lemma1} we have $j_1 \neq j_2$. Sets $\mathcal{P}_{(i,l_1)}$ and $\mathcal{P}_{(i,l_2)}$ are disjoint by definition. So from equation~(\ref{case1}) we have $B_{c(i,j_1)} \cap B_{c(i,j_2)} = \emptyset$.
\end{IEEEproof}

Note that our placement is partially flexible in nature (when $1 \leq t < \floor{b/z}$) in terms of which blocks are contained in $B_{c(i,j)}$.

\subsection{Delivery}\label{delivery}
In this subsection, we show both the delivery scheme and the correctness of the delivery scheme. 
\begin{lemma}\label{lemmad5}
User $k(i,j)$ where $1\leq i\leq m$, $1\leq j\leq b$ receives $(t^\prime(z-1) + t_z)b^{m-1}$ subfiles of each file $W^1,W^2,\ldots,W^N$ from the $z$ caches contained in $C_{k(i,j)}$. 
\end{lemma}
\begin{IEEEproof}
Let $C_{k(i,j)} = \{c(i,j_1), c(i,j_2), \ldots ,c(i,j_z)\}$. Due to condition \textbf{C2}, if $c(i,j_x) \in C_{(i,l_x)}$ and $c(i,j_y) \in C_{(i,l_y)}$ for $x,y \in \{1,2,\ldots,z\}$, $x \neq y$, then $C_{(i,l_x)} \neq C_{(i,l_y)}$. Hence, as shown in  Lemma~\ref{lemma2}, $B_{c(i,j_x)} \cap B_{c(i,j_y)} = \emptyset$. Hence, user $k(i,j)$ receives $(t^\prime (z-1))b^{m-1}$ subfiles of each file from $\{c(i,j_1), c(i,j_2), \ldots , c(i,j_{z-1})\}$ and an additional (non-overlapping) $t_zb^{m-1}$ subfiles of each file from $c(i,j_z)$ where $c(i,j_z) \in C_{(i,z)}$.
\end{IEEEproof}

\begin{lemma}\label{lemmad6}
When $t \geq b - (z-1)\floor{\frac{b}{z}}$, we have $R = 0$ files. 
\end{lemma}
\begin{IEEEproof}
In this case we have $t^\prime = \floor{b/z}$ and $t_z = b - (z-1)\floor{\frac{b}{z}}$. As per Lemma~\ref{lemmad5}, user $k(i,j)$ for $1\leq i\leq m$, $1\leq j\leq b$ receives $(t^\prime(z-1) + t_z)b^{m-1} = b^m$ subfiles of each file from the caches. As subpacketization level is $b^m$, every user receives all subfiles of all files at the server.
\end{IEEEproof}

%User $k(i,j)$ where $1\leq i\leq m$, $1\leq j\leq b$ can receive $t(z-1) + t^*$ subfiles of each file $W^1,W^2,\ldots,W^N$ from the $z$ caches contained in $C_{k(i,j)}$. 

In the remaining part of this subsection, we show the delivery when $t < b - (z-1)\floor{\frac{b}{z}}$. For each user $k(i,j)$ where $1\leq i\leq m$, $1\leq j\leq b$, we define a set $B_{k(i,j)} = \cup_{c(x,y) \in C_{k(i,j)}} B_{c(x,y)}$ (it is the set of blocks whose elements are the indices of the subfiles stored by the caches accessed by user $k(i,j)$).

For $1\leq i\leq m$, $1\leq j\leq b$, each user $k(i,j)$ needs to retrieve the subfiles indexed by the contents of the blocks in $\mathcal{P}_{i}\setminus B_{k(i,j)}$. For this purpose the users use the broadcast from the central server along with the contents of the accessible caches. The central server broadcasts following the below procedure.

%\subsubsection*{Broadcast strategy}
  %The definition is valid ev
%\begin{definition}
%\subsubsection*{Demand graph based delivery:}

We now construct a new graph that we call the demand graph. For the user-to-cache association bipartite graph $G = (U,C,E)$ and the MCRD $(X,\mathcal{A})$ based placement scheme (that is, $B_{c(i,j)}$ is defined for all $c(i,j) \in C$), the demand graph $\bar{G} = (V_1,V_2,\bar{E})$ is a bipartite graph  where $|V_1| = |V_2| = mb$, and for each cache in $C$ there is a unique vertex in $V_1$, for each block in $\mathcal{A}$ there is a unique vertex in $V_2$. In other words, there is a bijection from the vertices in $V_1$ to the caches in $C$, and there is another bijection from  the vertices in $V_2$ to the blocks in $\mathcal{A}$. Say for $l = 1,2$, $V_{l} = \{v_{(l,i,j)} | 1\leq i\leq m, 1\leq j\leq b\}$, and that the vertex $v_{(1,i,j)}$ maps to the cache $c(i,j)$, similarly, say that the vertex $v_{(2,i,j)}$ maps to the block $B(i,j)$.  There exists an edge in $\bar{E}$ connecting vertex $v_{(1,i,j)} \in V_1$ to vertex $v_{(2,\bar{i},j^*)} \in V_2$ if and only if $i = \bar{i}$ and $B(i,j^*) \notin B_{k(i,j^\prime)}$ where $k(i,j^\prime) = f_{M_i}^{-1}(c(i,j))$. The existence of $f_{M_i}$ and its inverse is ensured by condition \textbf{C3}.
%\end{definition}

%Construct a bipartite graph $\bar{G} = (V_1,V_2,\bar{E})$ where, $|V_1| = |V_2| = mb$, for each cache in $C$ ($C = \cup_{i=1}^m C_i$) there is a unique vertex in $V_1$, for each block in $\mathcal{A}$ there is a unique vertex in $V_2$. 

%\begin{example}

%\end{example}

We now show that conditions \textbf{C1} and \textbf{C2} enforces $(b - t^\prime (z-1) - t_z) b^m$ partial matchings (each containing $m$ edges) of a certain kind on $\bar{G}$. For $1\leq i\leq m$, $l = 1,2$ we define the set $V_{(l,i)} = \{v_{(l,i,j)} | 1\leq j\leq b \}$. Each vertex $v_{(1,i,j)} \in V_{(1,i)}$ is adjacent to $(b - t^\prime (z-1) - t_z)$ edges as there are $(b - t^\prime (z-1) - t_z)$ blocks in $\mathcal{P}_i$ that are not contained in $B_{f_{M_i}^{-1}(c(i,j))}$.% does not access.

To construct a partial matching of $m$ edges, pick $m$ vertices from $V_1$ by selecting only one vertex from each $V_{(1,i)}$ for $1\leq i\leq m$. There are $b$ ways to choose a vertex from $V_{(1,i)}$, and hence $b^m$ ways to choose these $m$ vertices from $V_1$. Consider one such chosen set to $m$ vertices. For each chosen vertex $v_{(1,i,j_i)} \in V_{(1,i)}$, pick one vertex $v \in V_{(2,i)}$ such that there is an edge connecting $v_{(1,i,j_i)}$ and $v$. It can be seen that $v$ can be chosen in $(b - t^\prime (z-1) - t_z)$ ways. 

For $1\leq i\leq m$, $1\leq j\leq b$, for each user $k(i,j)$ we define a bijection 
\begin{IEEEeqnarray*}{l}
f_{(i,j)}: \{1,2,\ldots, b - (t^\prime(z-1) + t_z)\} \to \{1,2,\ldots, b\} \setminus \{l \,|\, B(i,l) \in B_{k(i,j)}\}.
\end{IEEEeqnarray*}
The bijection $f_{(i,j)}$ can be defined arbitrarily without violating the domain, range, and bijectivity.% This map will be used to 
% Each user 
%
%
For $1\leq n\leq b - (t^\prime(z-1) + t_z)$ define a set of partial matchings $\mathcal{M}_n$ as following
\begin{IEEEeqnarray*}{l}
\mathcal{M}_n = \{ M^n_{j_1,j_2,\ldots,j_m} | 1\leq j_i \leq b, 1\leq i\leq m\}, \text{ where}\\
M^n_{j_1,j_2,\ldots,j_m} = \{(v_{(1,i,j_i)}, v_{(2,i,f_{(i,j_i^\prime)}(n))})\, | \, 1\leq i\leq m \} \text{ where }
k(i,j_i^\prime) = f_{M_i}^{-1}(c(i,j_i)).
\end{IEEEeqnarray*}
It can be seen that $\mathcal{M}_n$ has $b^m$ matchings. Let $\mathcal{M} = \cup_{n=1}^{b - (t^\prime(z-1) + t_z)} \mathcal{M}_n$. Hence, $\mathcal{M}$ has $(b - (t^\prime(z-1) + t_z))b^m$ matchings.
%

%Consider one such choice for each of the $m$ vertices chosen from $V_1$. Now, it can be seen that the set of $m$ edges adjacent to the  $m$ vertices chosen from $V_1$ forms a partial matching in $G = (V_1,V_2,\bar{E})$. Counting all the choices shows $(b - t^\prime (z-1) - t_z)^m b^m$ partial matchings each containing $m$ edges. Let $\mathcal{M}$
%denote the set of these $(b - t^\prime (z-1) - t_z)^m b^m$ partial matchings.
% be a subset of these $(b - t^\prime (z-1) - t_z)^m b^m$ partial matchings such that any edge $e \in \bar{E}$ appears in exactly one partial matching contained in $\mathcal{M}$. There could be more than one ways to construct $\mathcal{M}$, but it must be that $|\mathcal{M}| = (b - t^\prime (z-1) - t_z) b^m$.

%Let $\mathcal{M}_n \subseteq \mathcal{M}$ for $1\leq n \leq b - (t^\prime(z-1) + t_z)$ be a set of partial matchings defined as $\mathcal{M}_n = \{\mathcal{M}^1_n,\mathcal{M}^2_n, \ldots , \mathcal{M}^b_n\}$ where for $1\leq j\leq b$
%\begin{IEEEeqnarray*}{l}
%\mathcal{M}^j_n = \{(v_{(1,i,j)}, v_{2,i,f_{(i,j)}(n)})\, | \, 1\leq i\leq m \}
%\end{IEEEeqnarray*}
%It can be seen that $\mathcal{M}^j_n$ for $1\leq j\leq b$ is a partial matching having $m$ edges.  

% denote the set of all of these partial matchings.

During delivery, for each such partial matching in $\mathcal{M}$ we make one transmission, and each of these transmissions benefit all $m$ users associated with the matching (the association that is established by the inverse of $f_{M_i}$ acting on the $m$ chosen vertices (caches) from $V_1$). %Moreover, the delivery algorithm achieves a coding gain of $m$.% We emphasize that we do not know if matchings leads to coding gain, we state 
%M^n_{j_1,j_2,\ldots,j_m} | 1\leq j_i \leq b, 1\leq i\leq m
Say for a matching in $M^n_{j_1,j_2,\ldots,j_m} \in \mathcal{M}$ the vertex $v_{(1,i,j_i)} \in V_{(1,i)}$ connects to the vertex $v_{(2,i,j_i^*)} \in V_{(2,i)}$ for $1\leq i\leq m$. The transmission corresponding to this matching is given in the following. For $1\leq i\leq m$ compute the set $S^{n,i}_{j_1,j_2,\ldots,j_m}$ where
\begin{IEEEeqnarray}{l}
S^{n,i}_{j_1,j_2,\ldots,j_m} = B(1,j_1) \cap B(2,j_2) \cap \cdots \cap B(i-1,j_{i-1}) \cap B(i,j_i^*) \cap B(i+1,j_{i+1}) \cap \cdots \cap B(m,j_m).\label{eqn:bi:1}
\end{IEEEeqnarray}
Since in $(X,\mathcal{A})$ we have $\mu_m = 1$, set $S^{n,i}_{j_1,j_2,\ldots,j_m}$ for $1\leq i\leq m$ is a singleton set. We transmit the following $(b - (t^\prime(z-1) + t_z))b^m$ linear function of the subfiles to meet all users' demands.
%
%As $|S_i| = \mu_m$, say $S_i = \{y_{(i,1)},y_{(i,2)},\ldots,y_{(i,\mu_m)}\}$. For $1\leq l\leq mu_m$, transmit $Y_l$ where
\begin{IEEEeqnarray}{l}
Y^n_{j_1,j_2,\ldots,j_m} = \sum_{i=1}^m W^{d_{f_{M_i}^{-1}(c(i,j_i))}}(S^{n,i}_{j_1,j_2,\ldots,j_m}) \qquad \text{ for } 1\leq n\leq b - (t^\prime(z-1) + t_z), 1\leq j_1,j_2,\ldots,j_m \leq b. \IEEEeqnarraynumspace\label{eqn:bi:2}
\end{IEEEeqnarray}

\begin{lemma}\label{lemmad1}
For the set $S^{n,i}_{j_1,j_2,\ldots,j_m}$ defined in equation~(\ref{eqn:bi:1}), if $e \in S^{n,i}_{j_1,j_2,\ldots,j_m}$, then 
\newline \noindent (i) $e$ is not an element of any block contained in $B_{k(i,j_i^\prime)}$ where $k(i,j_i^\prime) = f_{M_i}^{-1}(c(i,j_i))$,
\newline \noindent (ii) $e$ is an element of some block in $B_{k(x,j_x^\prime)}$ where $k(x,j_x^\prime) = f_{M_x}^{-1}(c(x,j_x))$ for $1\leq x\leq m$, $x\neq i$,
\newline \noindent (iii) Using the transmission in equation~(\ref{eqn:bi:2}) user $k(i,j_i^\prime)$ retrieves the subfile $W^{d_{k(i,j_i^\prime)}}(S^{n,i}_{j_1,j_2,\ldots,j_m})$ where $f_{M_i}(k(i,j_i^\prime)) = c(i,j_i)$.
\end{lemma}
\begin{IEEEproof}
\noindent (i) Due to the construction of $\bar{G} = (V_1,V_2,\bar{E})$ we have $B(i,j_i^*) \notin B_{k(i,j_i^\prime)}$. 

%From equation~(\ref{eqn:bi:1}), we see that $e \in B(i,j_i^*)$. Say $c(i,j_i) \in C_{(i,l)}$ for some $1\leq l\leq z$. As per condition \textbf{C2}, user $k(i,j_i^\prime)$ does not access any cache in $C_{(i,l)}$ except $c(i,j_i)$. From equation~\ref{case1} it can be concluded that if a cache does not belong to $C_{(i,l)}$ then it does not store any subfile indexed by the blocks in $B_{c(i,j_i)}$. 

% Since the blocks of a parallel class are disjoint, $e$ is not present in any blocks contained in $B_{k(i,j_i)}$.

\noindent (ii) From equation~(\ref{eqn:bi:1}), we also see that $e \in B(x,j_x)$ for all $1\leq x\leq m$, $i \neq x$. Due to equation~(\ref{case1}), $B(x,j_x) \in B_{c(x,j_x)}$. We know user $k(x,j_x^\prime)$ accesses the cache $c(x,j_x)$. So $B(x,j_x) \in B_{k(x,j_x^\prime)}$.

\noindent (iii) As per the proof of statement (ii), user $k(i,j_i^\prime)$ knows any subfile indexed by the element contained in $S^{n,x}_{j_1,j_2,\ldots,j_m}$ for $1\leq x\leq m$, $x \neq i$. So in the sum  of equation~(\ref{eqn:bi:2}) user $k(i,j_i^\prime)$ knows $\sum_{x=1,x\neq i}^m W^{d_{k(x,j_x^\prime)}}(S^{n,x}_{j_1,j_2,\ldots,j_m})$. Hence user $k(i,j_i^\prime)$ can retrieve $W^{d_{k(i,j_i^\prime)}}(S^{n,i}_{j_1,j_2,\ldots,j_m})$.
\end{IEEEproof}

\begin{lemma}\label{lemmad4}
After making one transmission (given by equation~(\ref{eqn:bi:2})) for each partial matchings in $\mathcal{M}$, each user receives all subfiles of the file it demands.
\end{lemma}
\begin{IEEEproof}
Say for user $k(i,j_i^\prime)$ for $1\leq i\leq m$, $1\leq j_i^\prime\leq b$, $f_{M_i}(k(i,j_i^\prime)) = c(i,j_i)$. Let $B(i,j_i^*)$ be a block such that $B(i,j_i^*) \notin B_{k(i,j_i^\prime)}$. Then there exists an edge $e$ connecting vertices $v_{(1,i,j_i)} \in V_1$ and $v_{(2,i,j_i^*)} \in V_2$. So for some $1\leq n\leq b - (t^\prime(z-1) + t_z)$, $f_{(i,j_i^\prime)}(n) = j_i^*$. Let $\mathcal{M}_{(n,e)} \subseteq \mathcal{M}_{n}$ be the set of all matchings in $\mathcal{M}_{n}$ such that $M \in \mathcal{M}_{(n,e)}$ if and only if $e \in M$. It can be seen that $|\mathcal{M}_{(n,e)}| = b^{m-1}$. Say the vertices of $\mathcal{M}_{(n,e)}$ that belong to $V_1$ are $v_{(1,1,j_1)}, v_{(1,2,j_2)}, \ldots , v_{(1,m,j_m)}$.% $\{v_{1,i,j_i} \,| 1\leq i\leq m\}$. 

There are $b^{m-1}$ transmissions (given by equation~(\ref{eqn:bi:2})) for all matchings in $\mathcal{M}_{(n,e)}$. Hence, as per  statement (iii) of Lemma~\ref{lemmad1} and  equation~(\ref{eqn:bi:1}), user $k(i,j_i^\prime)$ can retrieve $W^{d_{k(i,j_i^\prime)}}(S^{n,i}_{j_1,j_2,\ldots,j_m})$ for $1\leq j_l\leq b$, $1\leq l\leq m$, $l \neq i$.

Corollary~\ref{coro1} showed that the block $B(i,j_i^*) = \cup_{1\leq j_l\leq b, 1\leq l\leq m, l \neq i} S^{n,i}_{j_1,j_2,\ldots,j_m}$. Hence, after the transmissions corresponding to all matchings in $\mathcal{M}_{n}$,  user $k(i,j_i^\prime)$ receives all subfiles whose indices belong to $B(i,j_i^*)$. 

Since $f_{(i,j_i^\prime)}$ is onto, after the transmissions corresponding to all matchings in $\mathcal{M}$, user $k(i,j_i^\prime)$ receives all subfiles whose indices belong to any block not contained in $B_{k(i,j_i^\prime)}$.
\end{IEEEproof}

As a result of Lemma~\ref{lemmad4}, and as subpacketization level is $b^m$, when $t < b - (z-1)\floor{\frac{b}{z}}$, a rate of $b - (t^\prime(z-1) + t_z)$ files is achievable.
%Write at the end.

\end{document}